\documentclass[journal]{IEEEtran}
\usepackage[caption=false,font=footnotesize,labelfont=rm,textfont=rm]{subfig}
\usepackage{amsmath,amsfonts}
\usepackage{algorithmic}
\usepackage{algorithm}
\usepackage{array}
\usepackage{caption}
\usepackage{stfloats}
\usepackage{url}
\usepackage{threeparttable}
\usepackage{ragged2e}
\usepackage{verbatim}
\usepackage{graphicx}
\usepackage{cite}
\usepackage{amssymb,amsthm,enumerate}
\usepackage{bm}
\usepackage[T1]{fontenc} 
\usepackage{mathtools}

\newtheorem{theorem}{Theorem}

\newtheorem{definition}{Definition}
\newtheorem{lemma}{Lemma}

\newtheorem{assumption}{Assumption}

\newtheorem{remark}{Remark}

\begin{document}

\title{Privacy-Preserving Push-Pull Method for Decentralized Optimization via State Decomposition}

\author{Huqiang~Cheng,
        Xiaofeng~Liao,~\IEEEmembership{Fellow,~IEEE,}
        Huaqing~Li,~\IEEEmembership{Senior~Member,~IEEE}
        and~You~Zhao
\thanks{The authors would like to thank Huan Gao, an Associate Professor with the School of Automation, Northwestern Polytechnical University, for his precious guidance and help in experimental validation. This work was supported in part by the National Key R\&D Program of China under Grant 2018AAA0100101, in part by the National Natural Science Foundation of China under Grant 61932006 and 61772434, in part by the Chongqing technology innovation and application development project under Grant cstc2020jscx-msxmX0156, and in part by the Natural Science Foundation of Chongqing under Grant CSTB2022NSCQ-MSX1217.}
\IEEEcompsocitemizethanks{\IEEEcompsocthanksitem H. Cheng, X. Liao, and Y. Zhao are with Key Laboratory of Dependable Services Computing in Cyber Physical Society-Ministry of Education, College of Computer Science, Chongqing University, Chongqing, China, 400044. E-mail: huqiangcheng@126.com; xfliao@cqu.edu.cn; Zhaoyou1991sdtz@163.com.(Corresponding author: Xiaofeng Liao.)
\IEEEcompsocthanksitem H. Li is with Chongqing Key Laboratory of Nonlinear Circuits and Intelligent Information Processing, College of Electronic and Information Engineering, Southwest University, Chongqing, China, 400715. E-mail: huaqingli@swu.edu.cn.\protect\\
}}
%


\maketitle

\begin{abstract}
Distributed optimization is manifesting great potential in multiple fields, e.g., machine learning, control, and resource allocation. Existing decentralized optimization algorithms require sharing explicit state information among the agents, which raises the risk of private information leakage. To ensure privacy security, combining information security mechanisms, such as differential privacy and homomorphic encryption, with traditional decentralized optimization algorithms is a commonly used means. However, this would either sacrifice optimization accuracy or incur heavy computational burden. To overcome these shortcomings, we develop a novel privacy-preserving decentralized optimization algorithm, called PPSD, that combines gradient tracking with a state decomposition mechanism. Specifically, each agent decomposes its state associated with the gradient into two substates. One substate is used for interaction with neighboring agents, and the other substate containing private information acts only on the first substate and thus is entirely agnostic to other agents. For the strongly convex and smooth objective functions, PPSD attains a $R$-linear convergence rate. Moreover, the algorithm can preserve the agents' private information from being leaked to honest-but-curious neighbors. Simulations further confirm the results.
\end{abstract}


\section{Introduction}
In decentralized optimization, a network of $n$ agents collaboratively solve
\begin{flalign}
\label{Eq:1} \underset{\mathbf{x}\in \mathbb{R}^d}{\min}\,\,F\left( \mathbf{x} \right) =\sum_{i=1}^n{f_i\left( \mathbf{x} \right)},\tag{1}
\end{flalign}
where the local objective function $f_i:\mathbb{R}^d\rightarrow \mathbb{R}$ is known only by agent $i$. Decentralized computing behavior of the problem \eqref{Eq:1} has led to its widespread use in different domains, including machine learning \cite{Anastasiia2021}, control \cite{Liu2016}, resource allocation \cite{Zhang2020a,YXu2017}, etc.

In recent years, a fairly rich set of gradient-based algorithms were developed to solve the problem. The initial work is the distributed gradient descent method (DGD) \cite{Nedic2009a}, which essentially combines average consensus with gradient descent using decay step size. It attains a convergence rate of order $\mathcal{O}( \ln k/\sqrt{k} )$ (resp. $\mathcal{O}\left( \ln k/k \right)$) for convex (resp. strongly convex) objective functions. On the basics of DGD, a well-known gradient tracking mechanism is designed by introducing an auxiliary variable to substitute the local gradient in DGD. The mechnism can significantly improve the convergence rate and optimization accuracy. Representative algorithms include AugDGM \cite{Xu2015}, AsynDGM \cite{Xu2017}, DIGing \cite{Qu2017,Nedic2017}, Push-DIGing \cite{Nedic2017}, ADD-Opt \cite{Xi2017b}, SONATA \cite{Sun2016}, ASY-SONATA \cite{Tian2020}, $\mathcal{A}\mathcal{B}$ \cite{Xin2018b}, TV-$\mathcal{A}\mathcal{B}$ \cite{Saadatniaki2020,Nedic2022} and Push-Pull \cite{Du2018,Zhang2019,Pu2021}, which can attain $R$-linear\footnote{An algorithm attains a $R$-linear convergence rate if there exists a constant $c>0$ such that the output sequence $\{ \mathbf{x}^k \}$ of the algorithm satisfies $\lVert \mathbf{x}^k-\mathbf{x}^{\ast} \rVert \le c\lambda ^k$ for any $k$, where $\lambda \in \left( 0,1 \right)$ and $\mathbf{x}^{\ast}$ is the optimal solution \cite{Nedic2017}.} convergence rate. Other related work includes \cite{Shi2014,Nedic2015,Zhang2020}, etc.

In decentralized optimization, all agents are required to collaboratively deal with the problem through local communication, which involves the information transmission between agents. This may lead to information leakage in the presence of an adversary in the network. However, all the algorithms mentioned above do not pay attention to privacy issues, which is undesirable in practical applications. By way of example, in the rendezvous problem, all the agents need to determine an optimal position without revealing their initial positions, which is of great importance in an attack environment. However, if the problem is solved directly using a gradient-based rendezvous algorithm, it may lead to the exposure of the agents' initial positions without employing appropriate privacy mechanisms \cite{Huang2015}. Another example is collaborative supervised learning, where the dynamics of the algorithm require a sharing of models or gradients among agents. Yet, the information may contain some sensitive data such as private paychecks, medical records, etc \cite{Yan2012}. Therefore, research on privacy-preserving mechanisms for decentralized optimization algorithms is essential and meaningful.

To address privacy security in decentralized optimization, some efforts on privacy-preserving algorithms have been made. There are two hot types of privacy-preserving algorithms. One is the differential-privacy based algorithms \cite{Huang2015,Wang2023,Ding2021,Chen2023a}, the other is partially homomorphic encryption based ones \cite{Lu2018,Zhang2018a,Zhang2018b}. However, these two types of algorithms have distinct drawbacks. The former requires a compromise between optimization accuracy and privacy level, while the latter incurs an expensive computational burden due to the presence of encryption and decryption operations. Moreover, as indicated in \cite{Yan2012} and \cite{Lou2018}, preserving private information can be achieved by combining projection steps or adding uncertainty to the step size. Nevertheless, the former requires a individual node to acquire advance information about the optimal solution, whereas the latter cannot protect against arbitrarily large variants of the gradient. Other related work includes \cite{Gade2018,Li2020a,Wang2022}. Note that these references only consider undirected networks. Gao et al. in \cite{Gao2023} proposed an efficient privacy-preserving algorithm over directed network, which showed that privacy can be preserved by harnessing optimization dynamics' robustness. However, the optimization accuracy is degraded with the injection of randomness into the dynamics.

For the study of privacy mechanisms, there have been some recent results in privacy-preserving multi-agent consensus algorithms \cite{Charalambous2019,Gao2018,Mo2017,Ruan2019,Wang2019,Chen2023,Gao2022}.
These algorithms are mainly used to ensure the security of agents' initial values against adversaries. Inspired by these studies, we develop a state decomposition based privacy-preserving algorithm in decentralized optimization to preserve the gradient information. Note that protecting the gradient is equivalent to protecting the objective function (including function types and function parameters) over the whole dynamics. Therefore, the work in this paper is more interesting and challenging than consensus algorithms that only consider protecting initial values of agents.

The primary contributions of this work are listed below.
\begin{enumerate}[1)]
\item{We develop a state decomposition based privacy-preserving decentralized optimization algorithm, termed as PPSD, for unbalanced digraphs. Specifically, each agent decomposes its auxiliary state associated with the gradient into two substates. Only one substate is used for interaction with neighboring agents, while the other substate acts only on the first substate and is thus entirely invisible to other agents. Moreover, at the initial iteration, we exploit the intrinsic robustness of decentralized optimization dynamics with the private information embedded in arbitrary mixing weights;}
\item{In contrast to the privacy notions in unobservability \cite{Alaeddini2017,Pequito2014} and opacity \cite{Lefebvre2020,Saboori2013,Ramasubramanian2019} based methods, which only consider protecting the exact value, our privacy definition is more stringent. Specifically, the private information (gradient) of each participating agent is kept secret if honest-but-curious agents make endless estimates of the gradient using accessible information;}
\item{We rigorously prove theoretically that PPSD not only achieves a $R$-linear convergence rate, but also effectively protects the private information of the participating agents. This is the most obvious difference with differential privacy based algorithms and homomorphic encryption based algorithms, where the former sacrifice optimization accuracy and the latter incur additional computational overhead. Simulations further validate the results.}
\end{enumerate}

\emph{Organization:} Section II reviews some basics, the push-pull algorithm, and privacy concerns. Section III formally introduces privacy-preserving algorithm. The convergence analysis and privacy analysis of PPSD are presented in Section IV and Section V, respectively. Numerical validation experiments are provided in Section VI. Lastly, we summarize the work in Section VII.

\emph{Notations:} Let the notations of the form $x$, $\mathbf{x}$, and $\mathbf{X}$ denote the scalar, vector, and matrix, respectively. $\mathbb{R}$ (resp. $\mathbb{R}^d$) is the set of (resp. $d$-dimensional) real numbers. $\mathbb{N}$ is the set of positive integers. Let $\mathbf{0}_d$ and $\mathbf{1}_d$ be the $d$-dimensional all-zero vector and all-one vector, respectively. Let $\mathbf{I}_d$ and $\mathbf{O}_d$ be $d\times d$-dimensional identity matrix and all-zero matrix, respectively. We use $\nabla f\left( \cdot \right)$ to denote the gradient. For any two sets $\mathcal{S}_1$ and $\mathcal{S}_2$, $\mathcal{S}_1\setminus \mathcal{S}_2$ represents that the elements belongs to $\mathcal{S}_1$ not to $\mathcal{S}_2$. Let $\left| \mathcal{S} \right|$ be the cardinality of $\mathcal{S}$. $\otimes$ and $\lVert \cdot \rVert$ denote the Kronecker product and the $\ell _2$-norm, respectively.

\section{Fundamentals}
We begin with a brief overview of some basics.

\subsection{Network Topology}
Consider a directed network $\mathcal{G}=\left( \mathcal{V}, \mathcal{E} \right)$, where $\mathcal{V}=\left\{ 1,\cdots ,N \right\}$ is the agent set, and $\mathcal{E}$ is the set of sequenced agent pairs $\left( j,i \right)$ implying that agent $i$ can send information to agent $j$. For convenience, assume $\left( i,i \right) \notin \mathcal{E}$ for every $i\in \mathcal{V}$. Define $\mathcal{N}_{i}^{\text{in}}=\left\{ j\left| \left( i, j \right) \in \mathcal{E} \right.  \right\}$ and $\mathcal{N}_{i}^{\text{out}}=\left\{ j\left| \left( j, i \right) \in \mathcal{E} \right. \right\}$ as the set of in-neighbors and out-neighbors of agent $i$, respectively. Let $\left| \mathcal{N}_{i}^{\text{in}} \right|$ and $\left| \mathcal{N}_{i}^{\text{out}} \right|$ be the cardinalities of $\mathcal{N}_{i}^{\text{in}}$ and $\mathcal{N}_{i}^{\text{out}}$, respectively.

\begin{assumption}\label{A1}
The directed network $\mathcal{G}=\left( \mathcal{V}, \mathcal{E} \right)$ is strongly connected, i.e., there exists at least one directed path between any two agents $i,j\in \mathcal{V}$.
\end{assumption}

\begin{definition}($\mu$-strongly convex)\label{D1}
There exists a constant $\mu>0$ enabling the differentiable function $f:\mathbb{R}^d\rightarrow \mathbb{R}$ to satisfy, for any $\mathbf{x}_1,\mathbf{x}_2\in \mathbb{R}^d$
\begin{flalign}
\nonumber \left( f\left( \mathbf{x}_1 \right) -f\left( \mathbf{x}_2 \right) \right) ^{\top}\left( \mathbf{x}_1-\mathbf{x}_2 \right) \ge \mu \lVert \mathbf{x}_1-\mathbf{x}_2 \rVert ^2,
\end{flalign}
then $f$ is $\mu$-strongly convex.
\end{definition}

\begin{definition}($L$-smooth)\label{D1}
There exists a constant $L>0$ enabling the differentiable function $f:\mathbb{R}^d\rightarrow \mathbb{R}$ to satisfy, for any $\mathbf{x}_1,\mathbf{x}_2\in \mathbb{R}^d$
\begin{flalign}
\nonumber \lVert \nabla f\left( \mathbf{x}_1 \right) -\nabla f\left( \mathbf{x}_2 \right) \rVert \le L\lVert \mathbf{x}_1-\mathbf{x}_2 \rVert,
\end{flalign}
then $\nabla f$ is $L$-smooth.
\end{definition}

\begin{assumption}\label{A2}
The global objective function $F$ is $\mu$-strongly convex, and the local gradient $\nabla f_i$ is $L_i$-smooth.
\end{assumption}

From Assumption \ref{A2}, the global gradient $\nabla F$ is $\bar{L}$-smooth wherein $\bar{L}=\sum\nolimits_{i=1}^n{L_i}$. Define $L=\max _{i\in \mathcal{V}}\left\{ L_i \right\}$. Moreover, it implies there exists a unique exact solution (denoted by $\mathbf{x}^{\ast}$) for the problem \eqref{Eq:1}.

\subsection{Push-Pull Method}
For the exploration of decentralized optimization, the Push-Pull method \cite{Du2018,Zhang2019,Pu2021} is a sophisticated protocol, see Algorithm 1. Each agent $i$ maintains two state variables: $\mathbf{x}_{i}^{k}$ and $\mathbf{y}_{i}^{k}$, and the private information to be protected for each agent $i$ is the function $f_i\left( \mathbf{x}_i \right)$. Note that to preserve $f_i\left( \mathbf{x}_i \right)$, it suffices to preserve the gradient function $\nabla f_i\left( \mathbf{x}_i \right)$.

\begin{algorithm}[htb]
    \renewcommand{\thealgorithm}{1}
	\caption{Push-Pull Method}
	\label{alg:1}
	\begin{algorithmic}
		\STATE \textbf{Initial setting:} Each agent $i\in \mathcal{V}$ sets $\mathbf{x}_{i}^{0}\in \mathbb{R}^d$ and $\mathbf{y}_{i}^{0}=\nabla f_i\left( \mathbf{x}_{i}^{0} \right)$. Set $k=0$.
        \STATE \textbf{Step 1:} Agent $i$ pushes the computed $\mathbf{x}_{i}^{k}$ to its out-neighbors $l\in \mathcal{N}_{i}^{\text{out}}$.
        \STATE \textbf{Step 2:} Agent $i$ pulls $\mathbf{x}_{j}^{k}$, $j\in \mathcal{N}_{i}^{\text{in}}$, and computes
        \begin{flalign}
        \label{Eq:2} \mathbf{x}_{i}^{k+1}=\sum_{j\in \mathcal{N}_{i}^{\text{in}}\cup \left\{ i \right\}}{r_{ij}\mathbf{x}_{j}^{k}}-\gamma \mathbf{y}_{i}^{k}, \tag{2}
        \end{flalign}
        where $r_{ij}=1/( | \mathcal{N}_{i}^{\text{in}} |+1 ) $ is the mixing weight.
        \STATE \textbf{Step 3:} After updating $\mathbf{x}_{i}$, agent $i$ pushes the computed $\mathbf{C}_{li}^{k}\mathbf{y}_{li}^{k}$ to its out-neighbors $l\in \mathcal{N}_{i}^{\text{out}}$, where $\mathbf{C}_{li}^{k}$ is the mixing weighted matrix.
        \STATE \textbf{Step 4:} Agent $i$ pulls $\mathbf{C}_{ij}^{k}\mathbf{y}_{ij}^{k}$, $j\in \mathcal{N}_{i}^{\text{in}}$, and computes
        \begin{flalign}
        \label{Eq:3} &\mathbf{y}_{i}^{k+1}=\sum_{j\in \mathcal{N}_{i}^{\text{in}}\cup \left\{ i \right\}}{c_{ij}\mathbf{x}_{j}^{k}}+\nabla f_i( \mathbf{x}_{i}^{k+1} ) -\nabla f_i( \mathbf{x}_{i}^{k} ), \tag{3}
        \end{flalign}
        where $c_{ij}=1/( | \mathcal{N}_{j}^{\text{out}} |+1 )$ is the mixing weights.
        \STATE \textbf{Step 5:} Set $k\gets k+1$, until a stopping criteria is satisfied.
    \end{algorithmic}
\end{algorithm}

Let Assumptions 1-2 hold. Some known efforts \cite{Zhang2019,Pu2021} have demonstrated that Algorithm 1 can achieve a convergence rate $\mathcal{O}( \lambda ^k )$ with $\lambda \in ( 0,1 )$ if the step size $\gamma$ is chosen to be small enough.

\subsection{Privacy Concern}
We introduce the attacker model and explain that conventional push-pull method cannot protect the privacy owing to the sharing of explicit information among agents.

\begin{definition}
In the network, there is a group of agents that correctly adhere to the dynamic update protocol (labeled as corrupt agents), that collude with each other, and that attempt to deduce the gradient functions of other agents (labeled as normal agents) using the information accessible to themselves.
\end{definition}

In decentralized optimization, such confidential information to be preserved is the gradient functions $\nabla f_i( \cdot )$. Recall the entire dynamics of the Push-Pull method. At $k=0$, agent $i$ sends $\mathbf{x}_{i}^{0}$ and $c_{ji}\mathbf{y}_{i}^{0}=c_{ji}\nabla f_i( \mathbf{x}_{i}^{0} )$ to its out-neighbors $j\in \mathcal{N}_{i}^{\text{out}}$, where $\mathbf{y}_{i}^{0}=\nabla f_i( \mathbf{x}_{i}^{0} )$. Let the number of agent $i$'s out-neighbors be available for $j$. Since $c_{ji}$ as $c_{ji}=1/( | \mathcal{N}_{i}^{\text{out}} |+1 )$, agent $j$ is capable of uniquely inferring $\nabla f_i$ at $\mathbf{x}_{i}^{0}$. Note that $| \mathcal{N}_{i}^{\text{out}} |\in \left\{ 1,2,\cdots ,n-1 \right\}$ and thus the value of $| \mathcal{N}_{i}^{\text{out}} |$ is easily accessible. Accordingly, the gradient information of an agent running the push-pull algorithm is trivially deduced by its neighbors. The similar arguments can be applied to other commonly used algorithms, which also suffer from the privacy leakage issues.

Our task is to propose an decentralized optimization algorithm that satisfies the following two requirements:
\begin{enumerate}[i)]
\item{Exact convergence: At the end of the algorithm execution, each agent converges to a consensus optimal solution $\mathbf{x}^{\ast}$, meaning achieving the global minimum $f\left( \mathbf{x}^{\ast} \right)$. Moreover, the convergence rate of the algorithm is not affected. That is to say, the push-pull algorithm can achieve a linear convergence rate, as can privacy-preserving push-pull algorithm.}
\item{Privacy preservation: During the entire dynamics of the algorithm, the privacy, i.e., the function information $f_i\left( \mathbf{x}_i \right)$, of each normal agent $i$ needs to be protected from honest-but-curious attacks. Note that the protected private information does not include corrupt agents' information.}
\end{enumerate}

\section{Privacy-Preserving Push-Pull Method}
We develop a privacy-preserving push-pull method by using state-decomposition mechanism, called PPSD, whose details are given in Algorithm 2. The main idea is for each agent to decompose its gradient state $\mathbf{y}_i$ into two substates $\mathbf{y}_{\alpha ,i}$ and $\mathbf{y}_{\beta ,i}$, as illustrated in Fig. 1. The substate $\mathbf{y}_{\alpha ,i}$ is used in the communication with other agents while $\mathbf{y}_{\beta ,i}$ is used only in the internal dynamics of agent $i$. In other words, the substate $\mathbf{y}_{\alpha ,i}$ acts as $\mathbf{y}_{i}$ of original graph, and the sub-state $\mathbf{y}_{\beta ,i}$ is hidden from agent $i$' neighbors. The mixing weights between agents and between substates are reported in TABLE I.

\begin{table*}[!ht]
 \centering
 \caption{Parameter setting}
 \begin{threeparttable}
 \begin{tabular}{|c|p{7.5cm}|p{7.5cm}|}\hline
 \textbf{Parameter}               & \textbf{Iteration} $k=0$                      & \textbf{Iterations} $k \ge 1$ \\ \hline
 $\mathbf{\Lambda }_{i}^{k}$            & $\mathbf{\Lambda }_{i}^{k}=\text{diag}\{ \gamma _{i}^{k}( 1 ) ,\cdots ,\gamma _{i}^{k}( d ) \}$, where $\gamma _{i}^{k}( 1 ) ,\cdots ,\gamma _{i}^{k}( d )$ are arbitrary constants selected from $\mathbb{R}$       & $\mathbf{\Lambda }_{i}^{k}=\gamma \mathbf{I}_d$, $\gamma>0$ \\ \hline
 $\mathbf{\Phi }_{i,\alpha}^{k}$  & $\mathbf{\Phi }_{i,\alpha}^{k}=\text{diag}\{ \alpha _{i}^{k}( 1 ) ,\cdots ,\alpha _{i}^{k}( d ) \}$, where $\alpha _{i}^{k}( 1 ) ,\cdots ,\alpha _{i}^{k}( d )$ are arbitrary constants selected from $\mathbb{R}$       & $\mathbf{\Phi }_{i,\alpha}^{k}=\alpha _{i}^{k}\mathbf{I}_d$, where $\alpha _{i}^{k}$ takes any value from $[ \eta ,1 ]$ \\ \hline
$\mathbf{\Phi }_{i,\beta}^{k}$    & $\mathbf{\Phi }_{i,\beta}^{k}=\text{diag}\{ \beta _{i}^{k}( 1 ) ,\cdots ,\beta _{i}^{k}( d ) \}$, where $\beta _{i}^{k}( 1 ) ,\cdots ,\beta _{i}^{k}( d )$ are arbitrary constants selected from $\mathbb{R}$        & $\mathbf{\Phi }_{i,\beta}^{k}=\beta _{i}^{k}\mathbf{I}_d$, where $\beta _{i}^{k}$ takes any value from $[ \eta ,1 ]$ \\ \hline
$\mathbf{R}_{ij}^{k}$             & $\mathbf{R}_{ij}^{k}=\text{diag}\{ r_{ij}^{k}( 1 ) ,\cdots ,r_{ij}^{k}( d ) \}$, where $r_{ij}^{k}( 1 ) ,\cdots ,r_{ij}^{k}( d )$ are arbitrary constants selected from $\mathbb{R}$ for $j\in \mathcal{N}_{i}^{\text{in}}\cup \{ i \}$                  & $\mathbf{R}_{ij}^{k}=r_{ij}^{k}\mathbf{I}_d$, where $r_{ij}^{k}$ takes any value from $[ \eta ,1 ]$ for $j\in \mathcal{N}_{i}^{\text{in}}\cup \{ i \}$, and $\sum\nolimits_{j\in \mathcal{N}_{i}^{\text{in}}\cup \{ i \}}{r_{ij}^{k}}=1$ \\ \hline
$\mathbf{A}_{ij}^{k}$             & $\mathbf{A}_{ij}^{k}=\text{diag}\{ a_{ij}^{k}( 1 ) ,\cdots ,a_{ij}^{k}( d ) \}$, where $a_{ij}^{k}( 1 ) ,\cdots ,a_{ij}^{k}( d )$ are arbitrary constants selected from $\mathbb{R}$ for $j\in \mathcal{N}_{i}^{\text{in}}\cup \{ i \}$                   & $\mathbf{A}_{ij}^{k}=a_{ij}^{k}\mathbf{I}_d$, where $a_{ij}^{k}$ takes any value from $[ \eta ,1 ]$ for $j\in \mathcal{N}_{i}^{\text{in}}\cup \{ i \}$, and $\sum\nolimits_{j\in \mathcal{N}_{i}^{\text{in}}\cup \{ i \}}{a_{ij}^{k}}=1$ \\ \hline
$\mathbf{C}_{ji}^{k}$             & $\mathbf{C}_{ji}^{k}=\text{diag}\{ c_{ji}^{k}( 1 ) ,\cdots ,c_{ji}^{k}( d ) \}$, where $c_{ji}^{k}( 1 ) ,\cdots ,c_{ji}^{k}( d )$ are arbitrary constants selected from $\mathbb{R}$ for $j\in \mathcal{N}_{i}^{\text{out}}$, and $\mathbf{C}_{ii}^{k}=\mathbf{I}_d-\sum\nolimits_{j\in \mathcal{N}_{i}^{\text{out}}}{\mathbf{C}_{ji}^{k}}-\mathbf{\Phi }_{i,\alpha}^{k}$ &
$\mathbf{C}_{ji}^{k}=c_{ji}^{k}\mathbf{I}_d$, where $c_{ji}^{k}$ takes any value from $[ \eta ,1 ]$ for $j\in \mathcal{N}_{i}^{\text{out}}\cup \{ i \}$, and $\sum\nolimits_{j\in \mathcal{N}_{i}^{\text{out}}\cup \{ i \}}{c_{ji}^{k}}+\alpha _{i}^{k}=1$ \\ \hline
 \end{tabular}
 \end{threeparttable}
 \centering
\end{table*}

\begin{figure}[htbp]
\centering
\includegraphics[width=6cm]{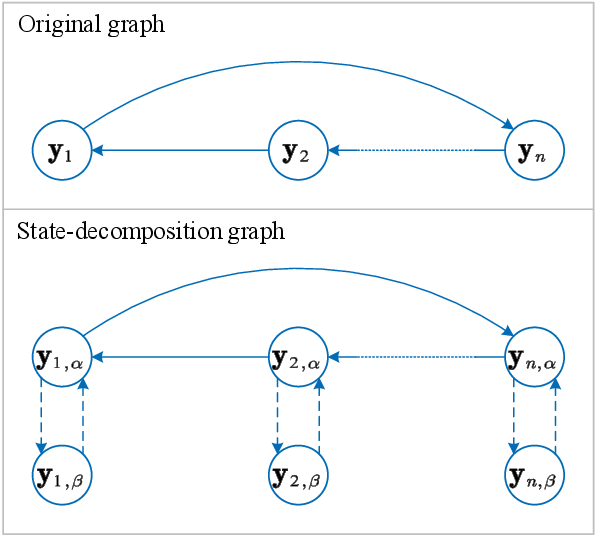}
\caption{State decomposition diagram.}
\end{figure}

\begin{algorithm}[htb]
	\caption{Privacy-Preserving Push-Pull Method}
	\label{alg:1}
	\begin{algorithmic}
		\STATE \textbf{Initial setting:} Each agent $i\in \mathcal{V}$ sets $\mathbf{x}_{i}^{0}\in \mathbb{R}^d$, $\mathbf{y}_{i,\alpha}^{0}\in \mathbb{R}^d$, and $\mathbf{y}_{i,\alpha}^{0}+\mathbf{y}_{i,\beta}^{0}=\nabla f_i\left( \mathbf{x}_{i}^{0} \right)$. At iteration $k$:
        \STATE \textbf{Step 1:} Agent $i$ pushes the computed $\mathbf{x}_{i}^{k}$ and $\mathbf{\Lambda }_{i}^{k}\mathbf{y}_{i,\alpha}^{k}$ to its out-neighbors $l\in \mathcal{N}_{i}^{\text{out}}$.
        \STATE \textbf{Step 2:} After pulling $\mathbf{x}_{j}^{k}$ and $\mathbf{\Lambda }_{j}^{k}\mathbf{y}_{j,\alpha}^{k}$ from $j\in \mathcal{N}_{i}^{\text{in}}$, agent $i$ computes
        \begin{flalign}
        \label{Eq:2} \mathbf{x}_{i}^{k+1}=\sum_{j\in \mathcal{N}_{i}^{\text{in}}\cup \left\{ i \right\}}{\mathbf{R}_{ij}^{k}\mathbf{x}_{j}^{k}-\mathbf{A}_{ij}^{k}\mathbf{\Lambda }_{j}^{k}\mathbf{y}_{j,\alpha}^{k}}. \tag{2}
        \end{flalign}
        \STATE \textbf{Step 3:} After updating $\mathbf{x}_{i}$, agent $i$ pushes the computed $\mathbf{C}_{li}^{k}\mathbf{y}_{li,\alpha}^{k}$ to $l\in \mathcal{N}_{i}^{\text{out}}$, where $\mathbf{C}_{li}^{k}$ is the mixing weighted matrix.
        \STATE \textbf{Step 4:} After pulling $\mathbf{C}_{ij}^{k}\mathbf{y}_{ij,\alpha}^{k}$ from $j\in \mathcal{N}_{i}^{\text{in}}$, agent $i$ computes
        \begin{flalign}
        \label{Eq:3} &\mathbf{y}_{i,\alpha}^{k+1}=\sum_{j\in \mathcal{N}_{i}^{\text{in}}\cup \left\{ i \right\}}{\mathbf{C}_{ij}^{k}\mathbf{y}_{j,\alpha}^{k}}+\left( \mathbf{I}_d-\mathbf{\Phi }_{i,\beta}^{k} \right) \mathbf{y}_{i,\beta}^{k}, \tag{3}
        \\
        \label{Eq:4} &\mathbf{y}_{i,\beta}^{k+1}\!=\!\mathbf{\Phi }_{i,\alpha}^{k}\mathbf{y}_{i,\alpha}^{k}\!+\!\mathbf{\Phi }_{i,\beta}^{k}\mathbf{y}_{i,\beta}^{k}\!+\!\nabla f_i\!\left( \mathbf{x}_{i}^{k\!+\!1} \right) \!-\!\nabla f_i\!\left( \mathbf{x}_{i}^{k} \right)\!, \tag{4}
        \end{flalign}
        where $\mathbf{\Phi }_{i,\alpha}^{k}$ and $\mathbf{\Phi }_{i,\beta}^{k}$ are diagonal matrices denoting internal sub-states' mixing weights.
        \STATE \textbf{Step 5:} Until a stopping criteria is satisfied, the dynamics stop, e.g., agent $i$ stops if $\lVert \mathbf{x}_{k}^{i}-\mathbf{x}^{\ast} \rVert<\epsilon$ for some predefined $\epsilon >0$.
    \end{algorithmic}
\end{algorithm}

\subsection{Privacy-preserving Policy}
To enable privacy, we let the private information $\nabla f_i( \mathbf{x}_{i}^{k} )$ be contained in $\mathbf{y}_{i,\beta}^{k}$ (shown in \eqref{Eq:4}). Since the dynamics of $\mathbf{y}_{i,\beta}^{k}$ occur only in the interior of the agent and the state $\mathbf{y}_{i,\beta}^{k}$ is not transmitted over the communication link, the private information $\nabla f_i( \mathbf{x}_{i}^{k} )$ is not disclosed. Moreover, albeit $\mathbf{y}_{i,\beta}^{k}$ will be used in the update of $\mathbf{y}_{i,\alpha}^{k}$ (shown in \eqref{Eq:3}), the randomness added in step size $\mathbf{\Lambda }_{i}^{k}$, mixing matrices $\mathbf{R}_{ij}^{k}$, $\mathbf{A}_{ij}^{k}$, and interior weights $\mathbf{\Phi }_{i,\alpha}^{k}$, $\mathbf{\Phi }_{i,\beta}^{k}$ at iteration $k=0$ contributes to preventing the leakage of $\nabla f_i( \mathbf{x}_{i}^{k} )$, see TABLE I. Specifically, at $k=0$, agent $i$ selects the mixing weights $\mathbf{C}_{ji}^{k}=\text{diag}\{ c_{ji}^{k}( 1 ) ,\cdots ,c_{ji}^{k}( d ) \}$ for $j\in \mathcal{N}_{i}^{\text{out}}$ arbitrarily in $\mathbb{R}$, which implies the weights may be negative. To guarantee the column stochasticity of $\mathbf{C}^k$, $\mathbf{C}_{ii}^{k}=\mathbf{I}_d-\sum\nolimits_{j\in \mathcal{N}_{i}^{\text{out}}}{\mathbf{C}_{ji}^{k}}-\mathbf{\Phi }_{i,\alpha}^{k}$. By a way, there is no need for $\mathbf{R}^k$ and $\mathbf{A}^k$ to be row-stochastic at iteration $k=0$. Thus, every agent $i$ can pick the weights $\{ \mathbf{R}_{ij}^{k} \}$ and $\{ \mathbf{A}_{ij}^{k} \}$ for $j\in \mathcal{N}_{i}^{\text{in}}\cup \{ i \}$ arbitrarily in the whole real number space. Further, as reported in TABLE I, the column stochasticity of $\mathbf{C}^k$ as well as the row stochasticity of $\mathbf{R}^k$ and $\mathbf{A}^k$ are necessary for $k\ge 1$.

Here, we use matrices $\mathbf{C}^k$ and $\mathbf{\Phi }_{\alpha}^{k}$ to explain how to ensure the settings in TABLE I are satisfied for $k\ge 1$. Specifically, every agent $i$ takes values from the set $\{ \alpha _{i}^{k}\in [ \eta ,1 ) ,c_{ji}^{k}\in [ \eta ,1 ) | i\in \mathcal{V}, j\in \mathcal{N}_{i}^{\text{out}}\cup \{ i \}  \}$ such that the sum of all elements is $1$, then set $\mathbf{\Phi }_{i,\alpha}^{k}$ and $\mathbf{C}_{ji}^{k}$ as $\mathbf{\Phi }_{i,\alpha}^{k}=\alpha _{i}^{k}\mathbf{I}_d$ and $\mathbf{C}_{ji}^{k}=c_{ji}^{k}\mathbf{I}_d$ for $i\in \mathcal{V}$ and $j\in \mathcal{N}_{i}^{\text{out}}\cup \{ i \}$, respectively. Note that such a set is easily obtained. As an instance, assume that each agent $i\in \mathcal{V}$ of original graph is decomposed into two sub-agents $i_{\alpha}$ and $i_{\beta}$. Sub-agent $i_{\alpha}$ takes the role of agent $i$ and sub-agent $i_{\beta}$ only communicates with sub-agent $i_{\alpha}$. Note that sub-agent $j_{\alpha}$ is an in/out-neighbor of sub-agent $i_{\alpha}$ if agent $j$ is an in/out-neighbor of agent $i$. Thus, we have $\mathcal{N}_{i_{\alpha}}^{\text{out}}=\{ j_{\alpha}| j\in \mathcal{N}_{i}^{\text{out}} \} \cup \{ i_{\beta} \}$ and $\mathcal{N}_{i_{\alpha}}^{\text{in}}=\{ j_{\alpha}| j\in \mathcal{N}_{i}^{\text{in}} \} \cup \{ i_{\beta} \}$. Agent $i_{\alpha}$ randomly selects a set of real values $\{ p_{j_{\alpha}i_{\alpha}}^{k}| j_{\alpha}\in \mathcal{N}_{i_{\alpha}}^{\text{out}}\cup \{ i_{\alpha} \} \}$ from $[ 0,1 ]$. Then, for $i\in \mathcal{V}$ and $j\in \mathcal{N}_{i}^{\text{out}}\cup \{ i \}$, we derive $c_{ji}^{k}$ and $\alpha _{i}^{k}$ by normalizing these values $\{ p_{j_{\alpha}i_{\alpha}}^{k} \}$ via
\begin{flalign}
\nonumber &c_{ji}^{k}=\frac{[ 1-( | \mathcal{N}_{i_{\alpha}}^{\text{out}} |+1 ) \eta ] [ ( 1-\eta ) p_{j_{\alpha}i_{\alpha}}^{k}+\eta ]}{( 1-\eta ) \sum\nolimits_{l_{\alpha}\in \mathcal{N}_{i_{\alpha}}^{\text{out}}\cup \{ i_{\alpha} \}}^{\,\,}{p_{l_{\alpha}i_{\alpha}}^{k}}+( | \mathcal{N}_{i_{\alpha}}^{\text{out}} |+1 ) \eta}+\eta,
\end{flalign}
\begin{flalign}
\nonumber &\alpha _{i}^{k}=\frac{[ 1-( | \mathcal{N}_{i_{\alpha}}^{\text{out}} |+1 ) \eta ] [ ( 1-\eta ) p_{i_{\beta}i_{\alpha}}^{k}+\eta ]}{( 1-\eta ) \sum\nolimits_{l_{\alpha}\in \mathcal{N}_{i_{\alpha}}^{\text{out}}\cup \{ i_{\alpha} \}}^{\,\,}{p_{l_{\alpha}i_{\alpha}}^{k}}+( | \mathcal{N}_{i_{\alpha}}^{\text{out}} |+1 ) \eta}+\eta.
\end{flalign}
One can verify that $\sum\nolimits_{j\in \mathcal{N}_{i}^{\text{out}}\cup \{ i \}}{c_{ji}^{k}}+\alpha _{i}^{k}=1$ and $c_{ji}^{k},\alpha _{i}^{k}\in [ \eta ,1 ]$ for $i\in \mathcal{V}$ and $j\in \mathcal{N}_{i}^{\text{out}}\cup \{ i \}$, and hence the column stochasticity of $\mathbf{C}^k$ is ensured by a decentralized manner.

\begin{remark}
Obviously, the state-decomposition graph has $2n$ agents, and it holds $| \mathcal{N}_{i_{\alpha}}^{\text{out}} |=| \mathcal{N}_{i}^{\text{out}} |+1$ and $| \mathcal{N}_{i_{\alpha}}^{\text{in}} |=| \mathcal{N}_{i}^{\text{in}} |+1$ for every $i\in \mathcal{V}$.
\end{remark}

Note that $\mathbf{R}_{ij}^{k}=\mathbf{A}_{ij}^{k}=\mathbf{C}_{ij}^{k}=\mathbf{O}_d$ for $j\notin \mathcal{N}_{i}^{\text{in}}\cup \left\{ i \right\}$. To facilitate the analysis, define
\begin{flalign}
\nonumber &\mathbf{\Phi }_{i,\alpha}^{k}=\text{diag}\left\{ \alpha _{i}^{k}\left( 1 \right) ,\cdots ,\alpha _{i}^{k}\left( d \right) \right\},
\\
\nonumber &\mathbf{\Phi }_{i,\beta}^{k}=\text{diag}\left\{ \beta _{i}^{k}\left( 1 \right) ,\cdots ,\beta _{i}^{k}\left( d \right) \right\},
\\
\nonumber &\mathbf{\Lambda }_{i}^{k}\!=\!\text{diag}\!\left\{ \gamma _{i}^{k}\left( 1 \right) ,\cdots ,\gamma _{i}^{k}\left( d \right) \right\}\!,\mathbf{x}^k\!=\!\left[ ( \mathbf{x}_{1}^{k} ) ^{\top},\cdots ,( \mathbf{x}_{n}^{k} ) ^{\top} \right]\! ^{\top},
\\
\nonumber &\mathbf{y}^k=[ ( \mathbf{y}_{1,\alpha}^{k} ) ^{\top},\cdots , ( \mathbf{y}_{n,\alpha}^{k} ) ^{\top}, ( \mathbf{y}_{1,\beta}^{k} ) ^{\top},\cdots , ( \mathbf{y}_{n,\beta}^{k} ) ^{\top} ] ^{\top},
\\
\nonumber &\nabla \mathbf{\hat{f}}( \mathbf{x}^k ) = [ \mathbf{0}_{nd}^{\top},( \nabla \mathbf{f}( \mathbf{x}^k ) ) ^{\top} ] ^{\top},
\\
\nonumber &\nabla \mathbf{f}( \mathbf{x}^k ) =[ ( \nabla f_1( \mathbf{x}_{1}^{k} ) ) ^{\top},\cdots ,( \nabla f_n( \mathbf{x}_{n}^{k} ) ) ^{\top} ] ^{\top}.
\end{flalign}
In addition, $\mathbf{R}_{ij}^{k}$, $\mathbf{A}_{ij}^{k}$, and $\mathbf{C}_{ij}^{k}$ are the $ij$-th block of $\mathbf{R}^k$, $\mathbf{A}^k$, and $\mathbf{C}^k$, respectively. $\mathbf{\Lambda }_{i}^{k}$, $\mathbf{\Phi }_{i,\alpha}^{k}$, and $\mathbf{\Phi }_{i,\beta}^{k}$ are the $i$-th diagonal block of $\mathbf{\Lambda }^k$, $\mathbf{\Phi }_{\alpha}^{k}$, and $\mathbf{\Phi }_{\beta}^{k}$, respectively. Then, we define
\begin{flalign}
\nonumber &\mathbf{\hat{C}}^k=\left[ \begin{matrix}
	\mathbf{C}^k&		\mathbf{I}_{nd}-\mathbf{\Phi }_{\beta}^{k}\\
	\mathbf{\Phi }_{\alpha}^{k}&		\mathbf{\Phi }_{\beta}^{k}\\
\end{matrix} \right], \mathbf{T}= \left[ \mathbf{I}_{nd}\,\,\mathbf{O}_{nd} \right].
\end{flalign}

Then, the dynamics \eqref{Eq:2}-\eqref{Eq:4} are equivalent to
\begin{flalign}
\label{Eq:5} &\mathbf{x}^{k+1}=\mathbf{R}^k\mathbf{x}^k-\mathbf{A}^k\mathbf{\Lambda }^k\mathbf{Ty}^k, \tag{5}
\\
\label{Eq:6} &\mathbf{y}^{k+1}=\mathbf{\hat{C}}^k\mathbf{y}^k+\nabla \mathbf{\hat{f}}\left( \mathbf{x}^{k+1} \right) -\nabla \mathbf{\hat{f}}\left( \mathbf{x}^k \right). \tag{6}
\end{flalign}
According to TABLE II, one can verify that $\mathbf{\hat{C}}^k$ is column-stochastic for every $k\ge 0$, while $\mathbf{R}^k$ is row-stochastic for $k \ge 1$.

\subsection{Relationship with Other Methods}
The proposed method has remarkably well generalization, and it can be transformed into many prominent algorithms by particular settings of parameters $\mathbf{R}^k$, $\mathbf{A}^k$, $\mathbf{\hat{C}}^k$, $\mathbf{\Lambda }^k$, and $\mathbf{T}$. As shown in TABLE II, With specific settings of the parameters, our algorithm can be transformed into existing algorithms. Also, it can derive some algorithms with new properties. Especially, settings such as the ones in this paper will lead to the algorithm with privacy-preserving capabilities.

\begin{table}[!ht]
 \centering
 \caption{Particular settings of parameters}
 \begin{threeparttable}
 \resizebox{\linewidth}{!}{
 \begin{tabular}{|c|c|c|c|c|c|}\hline
 \textbf{Algorithm}            & $\mathbf{R}^k$ & $\mathbf{A}^k$ & $\mathbf{\hat{C}}^k$ & $\mathbf{\Lambda}^k$ & $\mathbf{T}$ \\ \hline
 AsynDGM\cite{Xu2017}          & $\mathbf{W}^k$ & $\mathbf{W}^k$ & $\mathbf{W}^k$       & $\mathbf{\Lambda}$   & $\mathbf{I}$ \\ \hline
 DIGing\cite{Qu2017}           & $\mathbf{W}$   & $\mathbf{I}$   & $\mathbf{W}$         & $\lambda \mathbf{I}$ & $\mathbf{I}$ \\ \hline
 DIGing\cite{Nedic2017}        & $\mathbf{W}^k$ & $\mathbf{I}$   & $\mathbf{W}^k$       & $\lambda \mathbf{I}$ & $\mathbf{I}$ \\ \hline
 Push-Pull\cite{Du2018}        & $\mathbf{R}$   & $\mathbf{I}$   & $\mathbf{\bar{C}}$   & $\lambda \mathbf{I}$ & $\mathbf{I}$ \\ \hline
 Push-Pull\cite{Zhang2019}        & $\mathbf{R}$   & $\mathbf{R}$   & $\mathbf{\bar{C}}$   & $\lambda \mathbf{I}$ & $\mathbf{I}$ \\ \hline
 Push-Pull\cite{Pu2021}        & $\mathbf{R}$   & $\mathbf{R}$   & $\mathbf{\bar{C}}$   & $\mathbf{\Lambda}$   & $\mathbf{I}$ \\ \hline
 TV-$\mathcal{A}\mathcal{B}$\cite{Saadatniaki2020,Nedic2022}  & $\mathbf{R}^k$ & $\mathbf{I}$   & $\mathbf{\bar{C}}^k$ & $\lambda \mathbf{I}$ & $\mathbf{I}$ \\ \hline
 \end{tabular}}
 \justifying Here, $\mathbf{W}$, $\mathbf{R}$, and $\mathbf{\bar{C}}=\{c_{ij} \}$ (resp. $\mathbf{W}^k$, $\mathbf{R}^k$, and $\mathbf{\bar{C}}^k=\{ c_{ij}^{k} \}$) are time-invariant (resp. time-variant) doubly-stochastic, row-stochastic, and column-stochastic matrices, respectively. Moreover, $\lambda \mathbf{I}$ and $\mathbf{\Lambda}$ denote coordinated and uncoordinated step-size matrices, respectively.
 \end{threeparttable}
\end{table}

\section{Convergence Performance}
Inspired by \cite{Saadatniaki2020}, we analyze the convergence rate of PPSD. Due to the randomness of parameters in TABLE I at iteration $k=0$, we need to demonstrate that these randomnesses do not affect the gradient tracking property of the variable $\mathbf{y}$. Set $\tilde{n}\coloneqq 2n$.

\begin{lemma}
Let the sequence $\{ \mathbf{y}^k \} _{k\in \mathbb{N}}$ be generated by PPSD and the parameters satisfy TABLE I. Then, it holds $( \mathbf{1}_{\tilde{n}}^{\top}\otimes \mathbf{I}_d ) ( \mathbf{y}^k-\nabla \mathbf{\hat{f}}( \mathbf{x}^k ) ) =\mathbf{0}_d$ for $k\in \mathbb{N}$.
\begin{proof}
Using the column stochasticity of $\mathbf{\hat{C}}^k$ in \eqref{Eq:6} gives
\begin{flalign}
\nonumber  &( \mathbf{1}_{\tilde{n}}^{\top}\otimes \mathbf{I}_d ) \mathbf{y}^{k+1}
\\
\nonumber =&( \mathbf{1}_{\tilde{n}}^{\top}\otimes \mathbf{I}_d ) \mathbf{\hat{C}}^k\mathbf{y}^k+( \mathbf{1}_{\tilde{n}}^{\top}\otimes \mathbf{I}_d ) ( \nabla \mathbf{\hat{f}}( \mathbf{x}^{k+1} ) -\nabla \mathbf{\hat{f}}( \mathbf{x}^k ) )
\\
\nonumber =&( \mathbf{1}_{\tilde{n}}^{\top}\otimes \mathbf{I}_d ) \mathbf{y}^k+( \mathbf{1}_{\tilde{n}}^{\top}\otimes \mathbf{I}_d ) ( \nabla \mathbf{\hat{f}}( \mathbf{x}^{k+1} ) -\nabla \mathbf{\hat{f}}( \mathbf{x}^k ) ),
\end{flalign}
Then, we can derive that
\begin{flalign}
\nonumber  ( \mathbf{1}_{\tilde{n}}^{\top}\otimes \mathbf{I}_d ) ( \mathbf{y}^{k+1}-\nabla \mathbf{\hat{f}}( \mathbf{x}^{k+1} ) )=&( \mathbf{1}_{\tilde{n}}^{\top}\otimes \mathbf{I}_d ) ( \mathbf{y}^k-\nabla \mathbf{\hat{f}}( \mathbf{x}^k ) )
\\
\nonumber =&( \mathbf{1}_{\tilde{n}}^{\top}\otimes \mathbf{I}_d ) ( \mathbf{y}^0-\nabla \mathbf{\hat{f}}( \mathbf{x}^0 ) )
\end{flalign}
Recalling the initial setting of PPSD and the definition of the stack variable $\mathbf{y}^{k}$, one can easily verify that $( \mathbf{1}_{\tilde{n}}^{\top}\otimes \mathbf{I}_d ) \mathbf{y}^0=( \mathbf{1}_{\tilde{n}}^{\top}\otimes \mathbf{I}_d ) \nabla \mathbf{\hat{f}}( \mathbf{x}^0 )$. Combining the above relations yields the desired result.
\end{proof}
\end{lemma}

\begin{remark}
Lemma 1 indicates that the randomness of mixing weights at iteration $k=0$ has no influence on the gradient tracking performance of the variable $\mathbf{y}$. Moreover, the relation $( \mathbf{1}_{\tilde{n}}^{\top}\otimes \mathbf{I}_d ) \mathbf{y}^k=( \mathbf{1}_{\tilde{n}}^{\top}\otimes \mathbf{I}_d ) \nabla \mathbf{\tilde{f}}( \mathbf{x}^k ) =( \mathbf{1}_{n}^{\top}\otimes \mathbf{I}_d ) \nabla \mathbf{f}( \mathbf{x}^k )$ holds.
\end{remark}

Next, we analyze the system dynamics for $k>0$. From TABLE I, we construct some $n\times n$ matrices: $\mathbf{\bar{R}}^k=\{ r_{ij}^{k} \}$, $\mathbf{\bar{A}}^k=\{ a_{ij}^{k} \}$, $\mathbf{\bar{C}}^k=\{ c_{ij}^{k} \}$, $\mathbf{\bar{\Phi}}_{\alpha}^{k}=\text{diag}\{ \alpha _{1}^{k},\cdots ,\alpha _{n}^{k} \}$, $\mathbf{\bar{\Phi}}_{\beta}^{k}=\text{diag}\{ \beta _{1}^{k},\cdots ,\beta _{n}^{k} \}$, and the following matrices
\begin{flalign}
\nonumber \mathbf{\check{C}}^k=\left[ \begin{matrix}
	\mathbf{\bar{C}}^k&		\mathbf{I}_n-\mathbf{\bar{\Phi}}_{\beta}^{k}\\
	\mathbf{\bar{\Phi}}_{\alpha}^{k}&		\mathbf{\bar{\Phi}}_{\beta}^{k}\\
\end{matrix} \right] \in \mathbb{R}^{\tilde{n}\times \tilde{n}}, \,\, \mathbf{\bar{T}}=\left[ \mathbf{I}_n\,\,\mathbf{O}_n \right] \in \mathbb{R}^{n\times \tilde{n}}.
\end{flalign}
For $k\ge 1$, using these notions, the equations \eqref{Eq:5} and \eqref{Eq:6} can be reformulated as
\begin{flalign}
\label{Eq:7} &\mathbf{x}^{k+1}=( \mathbf{\bar{R}}^k\otimes \mathbf{I}_d ) \mathbf{x}^k-\gamma ( \mathbf{\bar{A}}^k\mathbf{\bar{T}}\otimes \mathbf{I}_d ) \mathbf{y}^k, \tag{7}
\\
\label{Eq:8} &\mathbf{y}^{k+1}=( \mathbf{\check{C}}^k\otimes \mathbf{I}_d ) \mathbf{y}^k+\nabla \mathbf{\hat{f}}( \mathbf{x}^{k+1} ) -\nabla \mathbf{\hat{f}}( \mathbf{x}^k ). \tag{8}
\end{flalign}
One can obtain that $\mathbf{R}^k=\mathbf{\bar{R}}^k\otimes \mathbf{I}_d$, $\mathbf{A}^k=\mathbf{\bar{A}}^k\otimes \mathbf{I}_d$, $\mathbf{C}^k=\mathbf{\bar{C}}^k\otimes \mathbf{I}_d$, $\mathbf{\Phi }_{\alpha}^{k}=\mathbf{\bar{\Phi}}_{\alpha}^{k}\otimes \mathbf{I}_d$, $\mathbf{\Phi }_{\beta}^{k}=\mathbf{\bar{\Phi}}_{\beta}^{k}\otimes \mathbf{I}_d$ for $k \ge 1$. For the sake of analysis, we introduce auxiliary variables: $\mathbf{s}^k=( ( \mathbf{\bar{V}}^k ) ^{-1}\otimes \mathbf{I}_d ) \mathbf{y}^k$ and $\mathbf{\bar{V}}^k=\text{diag}( \mathbf{v}^k )$ with
\begin{flalign}
\label{Eq:9} & \mathbf{v}^{k+1}=\mathbf{\check{C}}^k\mathbf{v}^k. \tag{9}
\end{flalign}
Let $\mathbf{v}^1=\frac{1}{\tilde{n}}\mathbf{1}_{\tilde{n}}$. Then, the equations \eqref{Eq:7} and \eqref{Eq:8} are transformed into
\begin{flalign}
\label{Eq:10} \mathbf{x}^{k+1}&\!=\!\mathbf{R}^k\mathbf{x}^k-\gamma ( \mathbf{\bar{A}}^k\mathbf{\bar{T}\bar{V}}^k\otimes \mathbf{I}_d ) \mathbf{s}^k, \tag{10}
\\
\label{Eq:11} \mathbf{s}^{k+1}&\!=\!\mathbf{P}^k\mathbf{s}^k\!+\!( ( \mathbf{\bar{V}}^{k+1} ) ^{-1}\!\otimes\! \mathbf{I}_d ) ( \nabla \mathbf{\hat{f}}( \mathbf{x}^{k+1} ) \!-\!\nabla \mathbf{\hat{f}}( \mathbf{x}^k ) ), \tag{11}
\end{flalign}
where $\mathbf{P}^k=\mathbf{\bar{P}}^k\otimes \mathbf{I}_d$ with $\mathbf{\bar{P}}^k=( \mathbf{\bar{V}}^{k+1} ) ^{-1}\mathbf{\check{C}}^k\mathbf{\bar{V}}^k$. One can verify from \cite{Saadatniaki2020,Gao2023} that $\mathbf{\bar{R}}^k$ and $\mathbf{\bar{P}}^k$ are row-stochastic and the sequences $\{ \mathbf{\bar{R}}^k \}$ and $\{ \mathbf{\bar{P}}^k \}$ are ergodic for $k\ge 1$. Note that $\{ \mathbf{v}^k \}$ is an absolute probability sequence of $\{ \mathbf{\bar{P}}^k \}$, and let $\{ \boldsymbol{\phi }^k \}$ be an absolute probability sequence of $\{ \mathbf{\bar{R}}^k \}$.

Lemmas 2 and 3 below show the contraction properties of $\mathbf{\bar{P}}^k$ and $\mathbf{\bar{R}}^k$, whose proofs follow similar arguments in \cite{Saadatniaki2020,Gao2023} and hence are omitted here.
\begin{lemma}
Suppose that Assumptions 1-2 hold. Let the parameters satisfy TABLE I. Set $Q_P\coloneqq 2\tilde{n}\frac{1+( \tilde{n}\eta ^{-\tilde{n}} ) ^{\tilde{n}-1}}{1-( \tilde{n}^{-1}\eta ^{\tilde{n}} ) ^{n-1}}$. Let $\mathbf{d}$ be an arbitrary vector in $\mathbb{R}^{\tilde{n}d}$, and $\mathbf{c}\coloneqq \mathbf{P}^k\mathbf{P}^{k-1}\cdots \mathbf{P}^{k-N_P+1}\mathbf{d}$. Then, for any $k\ge N_P$, it always holds
\begin{flalign}
\nonumber \lVert [ ( \mathbf{I}_{\tilde{n}}-\mathbf{1}_{\tilde{n}}( \mathbf{v}^{k+1} ) ^{\top} )& \otimes \mathbf{I}_d ] \mathbf{c} \rVert
\\
\nonumber \le &r_P\lVert [ ( \mathbf{I}_{\tilde{n}}-\mathbf{1}_{\tilde{n}}( \mathbf{v}^{k-N_P+1} ) ^{\top} ) \otimes \mathbf{I}_d ] \mathbf{d} \rVert.
\end{flalign}
where $N_P\in \mathbb{N}$ which makes $r_P\coloneqq Q_P( 1-( \tilde{n}^{-1}\eta ^{\tilde{n}} ) ^{\tilde{n}-1} ) ^{\frac{N_P-1}{\tilde{n}-1}}<1$ hold.
\end{lemma}

\begin{lemma}
Suppose that Assumptions 1-2 hold. Let the parameters satisfy TABLE I. Set $Q_R=2n\frac{1+\eta ^{-( n-1 )}}{1-\eta ^{n-1}}$. Let $\mathbf{d}$ be an arbitrary vector in $\mathbb{R}^{nd}$, and $\mathbf{c}=\mathbf{R}^k\mathbf{R}^{k-1}\cdots \mathbf{R}^{k-N_R+1}\mathbf{d}$. Then, for any $k\ge N_R$, it always holds
\begin{flalign}
\nonumber \lVert [ ( \mathbf{I}_n-\mathbf{1}_n( \boldsymbol{\phi }^{k+1} ) ^{\top} ) &\otimes \mathbf{I}_d ] \mathbf{c} \rVert
\\
\nonumber \le &r_R\lVert [ ( \mathbf{I}_n-\mathbf{1}_n ( \boldsymbol{\phi }^{k-N_R+1} ) ^{\top} ) \otimes \mathbf{I}_d ] \mathbf{d} \rVert,
\end{flalign}
where $N_R\in \mathbb{N}$ which makes $r_R=Q_R( 1-\eta ^{n-1} )^{\frac{N_R-1}{n-1}}<1$ hold.
\end{lemma}

Our proofs are based on the following quantities:
\begin{enumerate}[1)]
\item{Weighted average of $\mathbf{x}^k$: $\mathbf{\bar{x}}_{\text{w}}^{k}=( ( \boldsymbol{\phi }^k ) ^{\top}\otimes \mathbf{I}_d ) \mathbf{x}^k$;}
\item{Optimality gap: $\mathbf{r}^k=\mathbf{1}_n\otimes \mathbf{\bar{x}}_{\text{w}}^{k}-\mathbf{1}_n\otimes \mathbf{x}^{\ast}$;}
\item{Weighted consensus error: $\mathbf{\tilde{x}}_{\text{w}}^{k}=\mathbf{x}^k-\mathbf{1}_n\otimes \mathbf{\bar{x}}_{\text{w}}^{k}$;}
\item{Gradient estimation error: $\mathbf{\tilde{s}}_{\text{w}}^{k}=\mathbf{s}^k-( \mathbf{1}_{\tilde{n}}( \mathbf{v}^k ) ^{\top}\otimes \mathbf{I}_d ) \mathbf{s}^k$.}
\end{enumerate}
Next we provide some auxiliary results about the quantities defined above.
\begin{lemma}
Let the parameters satisfy TABLE I. Under Assumptions 1-2, it holds for $k \ge \bar{N}$:
\begin{flalign}
\nonumber &\lVert \mathbf{\tilde{x}}_{\text{w}}^{k+1} \rVert
\\
\nonumber \le &( r_R+\gamma Q_Rn\sqrt{n}L ) \lVert \mathbf{\tilde{x}}_{\text{w}}^{k-\bar{N}+1} \rVert +\gamma Q_Rn\sqrt{n}L\sum_{l=0}^{\bar{N}-2}{\lVert \mathbf{\tilde{x}}_{\mathbf{w}}^{k-l} \rVert}
\\
\nonumber            &+\gamma Q_Rn\sqrt{n}L\lVert \mathbf{r}^{k-\bar{N}+1} \rVert +\gamma Q_Rn\sqrt{n}L\sum_{l=0}^{\bar{N}-2}{\lVert \mathbf{r}^{k-l} \rVert}
\\
\label{Eq:12}          &+\gamma Q_R\sqrt{n}\lVert \mathbf{\tilde{s}}_{\text{w}}^{k-\bar{N}+1} \rVert +\gamma Q_R\sqrt{n}\sum_{l=0}^{\bar{N}-2}{\lVert \mathbf{\tilde{s}}_{\text{w}}^{k-l} \rVert}. \tag{12}
\end{flalign}
\begin{proof}
See Appendix A for proof.
\end{proof}
\end{lemma}

\begin{lemma}
Let the parameters satisfy TABLE I. Under Assumptions 1-2, it holds for $k \ge 1$:
\begin{flalign}
\label{Eq:13} \lVert \mathbf{r}^{k+1} \rVert \!\le\! \gamma nL\lVert \mathbf{\tilde{x}}_{\text{w}}^{k} \rVert \!+\!( 1\!-\!\gamma \tilde{n}^{-1}\eta ^{\tilde{n}-1}\mu )\! \lVert \mathbf{r}^k \rVert \!+\!\gamma n\lVert \mathbf{\tilde{s}}_{\text{w}}^{k} \rVert, \tag{13}
\end{flalign}
where $\gamma$ satisfies $0<\gamma \le 1/\bar{L}$.
\begin{proof}
See Appendix B for proof.
\end{proof}
\end{lemma}

\begin{lemma}
Let the parameters satisfy TABLE I. Under Assumptions 1-2, it holds for $k \ge \bar{N}$:
\begin{flalign}
\nonumber &\lVert \mathbf{\tilde{s}}_{\text{w}}^{k+1} \rVert
\\
\nonumber \le &\frac{( 2\tilde{n}\sqrt{n}LQ_P+\gamma n\tilde{n}\sqrt{n}L^2Q_P )}{\eta ^{\tilde{n}-1}}( \lVert \mathbf{\tilde{x}}_{\text{w}}^{k-\bar{N}+1} \rVert +\sum_{l=0}^{\bar{N}-2}{\lVert \mathbf{\tilde{x}}_{\text{w}}^{k-l} \rVert} )
\\
\nonumber &+\!( r_P\!+\!\frac{\gamma \tilde{n}\sqrt{n}LQ_P}{\eta ^{\tilde{n}-1}} ) \lVert \mathbf{\tilde{s}}_{\text{w}}^{k-\bar{N}+1} \rVert \!+\!\frac{\gamma \tilde{n}\sqrt{n}LQ_P}{\eta ^{\tilde{n}-1}}\sum_{l=0}^{\bar{N}-2}{\lVert \mathbf{\tilde{s}}_{\text{w}}^{k-l} \rVert}
\\
\label{Eq:14} &+\frac{\gamma n\tilde{n}\sqrt{n}L^2Q_P}{\eta ^{\tilde{n}-1}}( \lVert \mathbf{r}^{k-\bar{N}+1} \rVert +\sum_{l=0}^{\bar{N}-2}{\lVert \mathbf{r}^{k-l} \rVert} ), \tag{14}
\end{flalign}
\begin{proof}
See Appendix C for proof.
\end{proof}
\end{lemma}

Next, we present the main convergence result of PPSD. For ease of expression, we define
\begin{flalign}
\nonumber &\mathbf{U}_a=\left[ \begin{matrix}
	\gamma nLq_2&		\gamma nLq_2&		\gamma q_2\\
	\gamma nL&		1-\gamma \eta ^{n-1}q_3&		\gamma n\\
	2q_1+\gamma nLq_1&		\gamma nLq_1&		\gamma q_1\\
\end{matrix} \right] ,
\\
\nonumber &\mathbf{U}_b=\left[ \begin{matrix}
	\gamma nLq_2&		\gamma nLq_2&		\gamma q_2\\
	0&		0&		0\\
	2q_1+\gamma nLq_1&		\gamma nLq_1&		\gamma q_1\\
\end{matrix} \right] ,
\\
\nonumber &\mathbf{U}_c=\left[ \begin{matrix}
	r_R+\gamma nLq_2&		\gamma nLq_2&		\gamma q_2\\
	0&		0&		0\\
	2q_1+\gamma nLq_1&		\gamma nLq_1&		r_P+\gamma q_1\\
\end{matrix} \right],
\end{flalign}
where $q_1=\frac{\tilde{n}\sqrt{n}LQ_P}{\eta ^{\tilde{n}-1}}$, $q_2=Q_R\sqrt{n}$, and $q_3=n^{-1}\mu$.

\begin{theorem}
Let the parameters satisfy TABLE I. Under Assumptions 1-2, PPSD achieve a $R$-linear rate if $\gamma$ is chosen to be small enough.
\begin{proof}
Using the results in Lemmas 2,3, and 4 gives
\begin{flalign}
\nonumber \left[ \begin{array}{c}
	\boldsymbol{\zeta }^{k+1}\\
	\boldsymbol{\zeta }^k\\
	\boldsymbol{\zeta }^{k-1}\\
	\vdots\\
	\boldsymbol{\zeta }^{k-\bar{N}+2}\\
\end{array} \right] \le \underset{\triangleq \mathbf{U}\left( \gamma \right)}{\underbrace{\left[ \begin{matrix}
	\mathbf{U}_a&		\mathbf{U}_b&		\cdots&		\mathbf{U}_b&		\mathbf{U}_c\\
	\mathbf{I}_3&		\,\,&		\,\,&		\,\,&		\,\,\\
	\,\,&		\mathbf{I}_3&		\,\,&		\,\,&		\,\,\\
	\,\,&		\,\,&		\ddots&		\,\,&		\,\,\\
	\,\,&		\,\,&		\,\,&		\mathbf{I}_3&		\,\,\\
\end{matrix} \right] }}\left[ \begin{array}{c}
	\boldsymbol{\zeta }^k\\
	\boldsymbol{\zeta }^{k-1}\\
	\vdots\\
	\boldsymbol{\zeta }^{k-\bar{N}+2}\\
	\boldsymbol{\zeta }^{k-\bar{N}+1}\\
\end{array} \right].
\end{flalign}
Further, the system matrix $\mathbf{U}( \gamma )$ can be partitioned as $\mathbf{U}^1+\gamma \mathbf{U}^2$, where
\begin{flalign}
\setlength{\arraycolsep}{1.5pt}
\nonumber \mathbf{U}^1=\left[ \begin{matrix}
	\mathbf{U}_{a}^{1}&		\mathbf{U}_{b}^{1}&		\cdots&		\mathbf{U}_{b}^{1}&		\mathbf{U}_{c}^{1}\\
	\mathbf{I}_3&		\,\,&		\,\,&		\,\,&		\,\,\\
	\,\,&		\mathbf{I}_3&		\,\,&		\,\,&		\,\,\\
	\,\,&		\,\,&		\ddots&		\,\,&		\,\,\\
	\,\,&		\,\,&		\,\,&		\mathbf{I}_3&		\,\,\\
\end{matrix} \right],
\mathbf{U}^2=\left[ \begin{matrix}
	\mathbf{U}_{a}^{2}&		\mathbf{U}_{b}^{2}&		\cdots&		\mathbf{U}_{b}^{2}&		\mathbf{U}_{c}^{2}\\
	\mathbf{O}_{3}&		\,\,&		\,\,&		\,\,&		\,\,\\
	\,\,&		\mathbf{O}_{3}&		\,\,&		\,\,&		\,\,\\
	\,\,&		\,\,&		\ddots&		\,\,&		\,\,\\
	\,\,&		\,\,&		\,\,&		\mathbf{O}_{3}&		\,\,\\
\end{matrix} \right]
\end{flalign}
with
\begin{flalign}
\nonumber &\mathbf{U}_{a}^{1}=\left[ \begin{matrix}
	0&		0&		0\\
	0&		1&		0\\
	2q_1&		0&		0\\
\end{matrix} \right],\,\,\mathbf{U}_{a}^{2}=\left[ \begin{matrix}
	nLq_2&		nLq_2&		q_2\\
	nL&		-\eta ^{n-1}q_3&		n\\
	nLq_1&		nLq_1&		q_1\\
\end{matrix} \right],
\\
\nonumber &\mathbf{U}_{b}^{1}=\left[ \begin{matrix}
	0&		0&		0\\
	0&		0&		0\\
	2q_1&		0&		0\\
\end{matrix} \right],\,\,\mathbf{U}_{b}^{2}=\left[ \begin{matrix}
	nLq_2&		nLq_2&		q_2\\
	0&		0&		0\\
	nLq_1&		nLq_1&		q_1\\
\end{matrix} \right],
\\
\nonumber &\mathbf{U}_{c}^{1}=\left[ \begin{matrix}
	r_R&		0&		0\\
	0&		0&		0\\
	2q_1&		0&		r_P\\
\end{matrix} \right],\,\,\mathbf{U}_{c}^{2}=\left[ \begin{matrix}
	nLq_2&		nLq_2&		q_2\\
	0&		0&		0\\
	nLq_1&		nLq_1&		q_1\\
\end{matrix} \right],
\end{flalign}
The key to achieving $R$-linear convergence in PPSD is to show that the spectral radius of $\mathbf{U}( \gamma )$ satisfies $\rho ( \mathbf{U}( \gamma ) ) <1$. First, from Theorem 3.2 in \cite{Powell2011}, one can verify that
\begin{flalign}
\nonumber \det ( \lambda \mathbf{I}-\mathbf{U}^1 ) =( \lambda ^{\bar{N}}-r_R ) ( \lambda ^{\bar{N}}-r_P ) ( \lambda -1 ) \lambda ^{\bar{N}-1}.
\end{flalign}
Due to $r_R,r_P\in ( 0,1 )$, it follows $\rho ( \mathbf{U}^1 ) =1$. Besides, $\mathbf{u}=\mathbf{1}_{\bar{N}}\otimes [ 0\,\, 1\,\, 0 ] ^{\top}$ (resp. $\mathbf{w}=[ 0\,\, 1\,\, 0 \cdots \,\,0 ] ^{\top}$) is the left (resp. right) eigenvector corresponding to the eigenvalue $1$ of $\mathbf{U}( \gamma )$. We treat the simple eigenvalue $p( \gamma )$ of $\mathbf{U}( \gamma )$ as a function about $\gamma$. Since $\mathbf{U}( \gamma ) =\mathbf{U}^1+\gamma \mathbf{U}^2$ holds, we have $p( 0 ) =1$. Applying Theorem 6.3.12 in \cite{Horn2012} gives
\begin{flalign}
\nonumber \left. \frac{dp( \gamma )}{d\gamma} \right|_{\gamma =0}=\frac{\mathbf{w}^{\top}\mathbf{U}^2\mathbf{u}}{\mathbf{w}^{\top}\mathbf{u}}=-n^{-1}\eta ^{n-1}\mu <0,
\end{flalign}
where $\mathbf{w}^{\top}\mathbf{u}=1$ and $\mathbf{w}^{\top}\mathbf{U}^2\mathbf{u}=-n^{-1}\eta ^{n-1}\mu$.

Since $p( \gamma )$ is a continuous function about $\gamma$, it holds $p( \gamma )<1$ if $\gamma$ is small enough, which means that $\rho ( \mathbf{U}( \gamma ) ) <1$. Moreover, all entries of $\mathbf{U}( \gamma )$ are nonnegative, and all entries of $\mathbf{U}( \gamma ) ^{\bar{N}+1}$ are positive. It is verified from Theorems 8.5.1 and 8.5.2 in \cite{Horn2012} that $\mathbf{U}( \gamma ) ^k$ converges to $0$ at least at the rate of $\mathcal{O}( \rho ( \mathbf{U}( \gamma ) ) ^k )$. Therefore, $\lVert \mathbf{x}^k-\mathbf{1}_n\otimes \mathbf{x}^{\ast} \rVert=\mathcal{O}( \rho ( \mathbf{U}( \gamma ) ) ^k )$ as long as $\gamma$ is sufficiently, implying PPSD can achieve $R$-linear convergence.
\end{proof}
\end{theorem}

\begin{remark}
Theorem 1 indicates that PPSD has $R$-linear convergence. In other words, the state-decomposition mechanism and arbitrary setting of parameters at iteration $k=0$ do not affect the convergence performance of PPSD.
\end{remark}

\section{Privacy Performance}
We prove that PPSD can preserve the private information of the normal agents against the honest-but-curious attacks.

Recalling Definition 3, we consider a set of corrupt agents $\mathcal{A}$ attempting to deduce the gradient information of normal agent $i$ using their accessible information. From PPSD, the information accessed by $\mathcal{A}$ at iteration $k$ is
\begin{flalign}
\nonumber \mathcal{I}_{\mathcal{A}}\left( k \right) =\left\{ \mathcal{I}_j\left( k \right) \left| j\in \mathcal{A} \right. \right\},
\end{flalign}
where $\mathcal{I}_j\left( k \right)$ denotes the information accessible to agent $j$ at iteration $k$. The mathematical form of $\mathcal{I}_j\left( k \right)$ is given as
\begin{flalign}
\nonumber &\mathcal{I}_j\left( k \right) =\left\{ \mathcal{I}_{j}^{\text{state}}\left( k \right) \cup \mathcal{I}_{j}^{\text{send}}\left( k \right) \cup \mathcal{I}_{j}^{\text{receive}}\left( k \right) \right\}
\\
\nonumber            &\cup \left\{ \mathbf{\Lambda }_{j}^{k},\mathbf{R}_{jj}^{k},\mathbf{A}_{jj}^{k},\mathbf{C}_{jj}^{k},\mathbf{\Phi }_{j,\alpha}^{k},\mathbf{\Phi }_{j,\beta}^{k} \right\}
\\
\nonumber            &\cup \left\{ \mathbf{R}_{jl}^{k},\mathbf{A}_{jl}^{k}\left| \forall l\in \mathcal{N}_{j}^{\text{in}} \right. \right\} \cup \left\{ \mathbf{C}_{mj}^{k}\left| \forall m\in \mathcal{N}_{j}^{\text{out}} \right. \right\}
\\
\nonumber            &\cup \left\{ \mathbf{\Lambda }_{l}^{k},\mathbf{R}_{lm}^{k},\mathbf{A}_{lm}^{k},\mathbf{C}_{lm}^{k},\mathbf{\Phi }_{l,\alpha}^{k}\left| \text{if}\,\, k\ge 1\,\, \text{and}\,\, \forall l,m\in \mathcal{V}\setminus \left\{ j \right\} \right. \right\}
\end{flalign}
where
\begin{flalign}
\nonumber &\mathcal{I}_{j}^{\text{state}}\left( k \right) =\left\{ \mathbf{x}_{j}^{k},\mathbf{y}_{j,\alpha}^{k},\mathbf{\Lambda }_{j}^{k}\mathbf{y}_{j,\alpha}^{k},\mathbf{C}_{jj}^{k}\mathbf{y}_{j,\alpha}^{k} \right\},
\\
\nonumber &\mathcal{I}_{j}^{\text{send}}\left( k \right) =\left\{ \mathbf{x}_{j}^{k},\mathbf{\Lambda }_{j}^{k}\mathbf{y}_{j,\alpha}^{k},\mathbf{C}_{mj}^{k}\mathbf{y}_{j,\alpha}^{k}\left| \forall m\in \mathcal{N}_{j}^{\text{out}} \right. \right\},
\\
\nonumber &\mathcal{I}_{j}^{\text{receive}}\left( k \right) =\left\{ \mathbf{x}_{l}^{k},\mathbf{\Lambda }_{l}^{k}\mathbf{y}_{l,\alpha}^{k},\mathbf{C}_{jl}^{k}\mathbf{y}_{l,\alpha}^{k}\left| \forall l\in \mathcal{N}_{j}^{\text{in}} \right. \right\}.
\end{flalign}

Give a constant $\kappa \in \mathbb{N}$, we let the information accessible to the $\mathcal{A}$ from $0$ to $\kappa$ be denoted as $\mathcal{I}_{\mathcal{A}}( 0:\kappa ) =\cup _{0\le k\le \kappa}\mathcal{I}_{\mathcal{A}}( k )$. For any $\mathcal{I}_{\mathcal{A}}( 0:\kappa )$, let $\nabla _i( \mathcal{I}_{\mathcal{A}}( 0:\kappa ) )$ be the set of all gradients $\nabla f_i( \cdot )$ at agent $i$ that make the information sequences generated by PPSD the same as the accessible ones to $\mathcal{A}$. That is, $\nabla _i( \mathcal{I}_{\mathcal{A}}( 0:\kappa ) )$ contains all potential gradients for which agent $i$  is capable of generating $\mathcal{I}_{\mathcal{A}}( 0:\kappa )$. Denote the diameter of $\nabla _i( \mathcal{I}_{\mathcal{A}}( 0:\kappa ) )$ by
\begin{flalign}
\nonumber \mathcal{D}_i( \mathcal{I}_{\mathcal{A}}( 0:\kappa ) ) \coloneqq \underset{\nabla f_i( \cdot ) ,\nabla \tilde{f}_i( \cdot ) \in \nabla _i( \mathcal{I}_{\mathcal{A}}( 0:\kappa ) )}{\text{sup}}\lVert \nabla f_i( \cdot ) -\nabla \tilde{f}_i( \cdot ) \rVert.
\end{flalign}

\begin{definition}
Consider a decentralized network of $n$ agents. There exists a set of corrupt agents, who can collude with each other. The private information of each normal agent $i$ is protected against corrupt agents $\mathcal{A}$ if $\mathcal{D}_i( \mathcal{I}_{\mathcal{A}}( 0:\kappa ) ) =\infty$ for every $\kappa \in \mathbb{N}$ and $\mathcal{I}_{\mathcal{A}}( 0:\kappa )$.
\end{definition}

The above definition is inspired from the notion of $l$-diversity, where the diversity of the private information is assessed by the number of different estimations for the information, and a greater diversity means more uncertain estimates of private information. In this work, we treat the gradient as the private information, and its confidentiality is assessed by $\mathcal{D}_i( \mathcal{I}_{\mathcal{A}}( 0:\kappa ) )$. The larger $\mathcal{D}_i( \mathcal{I}_{\mathcal{A}}( 0:\kappa ) )$, the greater the confidentiality.

\begin{remark}
Note that $\mathcal{D}_i( \mathcal{I}_{\mathcal{A}}( 0:\kappa ) ) =\infty$ implies achieving the largest possible uncertainty. Thus, the private information is preserved if $\mathcal{D}_i( \mathcal{I}_{\mathcal{A}}( 0:\kappa ) ) =\infty$. Our privacy definition is more stringent than the ones in unobservability based methods \cite{Alaeddini2017,Pequito2014} and opacity based methods \cite{Lefebvre2020,Saboori2013,Ramasubramanian2019}. Specifically, Definition 4 specifies that an adversary cannot find an exact value or even an effective range of $\nabla f_i( \cdot )$, while unobservability opacity and based methods only consider consider protecting the exact value.
\end{remark}

Based on the above discussion, we present the privacy performance of PPSD.
\begin{theorem}
Let the parameters satisfy TABLE I. In PPSD, the gradient information of each normal agent $i$ is preserved against corrupt agents $\mathcal{A}$ if $\mathcal{N}_{i}^{\text{out}}\cup \mathcal{N}_{i}^{\text{in}}\not\subset \mathcal{A}$.
\begin{proof}
Given $\kappa \in \mathbb{N}$, our task is to prove $\mathcal{D}_i( \mathcal{I}_{\mathcal{A}}( 0:\kappa ) ) =\infty$. Let $\mathcal{I}_{\mathcal{A}}( 0:\kappa )$ and $\tilde{\mathcal{I}}_{\mathcal{A}}( 0:\kappa )$ be the information generated under $\nabla f_i( \cdot )$ and $\nabla \tilde{f}_i( \cdot ) \coloneqq \nabla f_i( \cdot ) +\boldsymbol{\delta }$, respectively, with $\boldsymbol{\delta }=[ \boldsymbol{\delta }( 1 ) ,\cdots ,\boldsymbol{\delta }( d ) ] ^{\top}$ being an arbitrary vector in $\mathbb{R}^d$. The main idea is to make $\nabla f_i( \cdot ), \nabla \tilde{f}_i( \cdot ) \in \nabla _i( \mathcal{I}_{\mathcal{A}}( 0:\kappa ) )$ hold, i.e., $\mathcal{I}_{\mathcal{A}}( 0:\kappa )=\tilde{\mathcal{I}}_{\mathcal{A}}( 0:\kappa )$. Since $\mathcal{N}_{i}^{\text{out}}\cup \mathcal{N}_{i}^{\text{in}}\not\subset \mathcal{A}$, there exists an agent $m\in \mathcal{N}_{i}^{\text{out}}\cup \mathcal{N}_{i}^{\text{in}}\setminus \mathcal{A}$. Therefore, we only need to prove $\mathcal{I}_{\mathcal{A}}( 0:\kappa )=\tilde{\mathcal{I}}_{\mathcal{A}}( 0:\kappa )$ under the settings: $\nabla \tilde{f}_i( \cdot ) =\nabla f_i( \cdot ) +\boldsymbol{\delta }$, $\nabla \tilde{f}_m( \cdot ) =\nabla f_m( \cdot ) -\boldsymbol{\delta }$, and $\nabla \tilde{f}_l( \cdot ) =\nabla f_l( \cdot )$ for $l\in \mathcal{V}\setminus \{ i,m \}$. Moreover, to ensure the initial conditions of PPSD, the variables' initial values are set as follows:
\begin{flalign}
\nonumber &\mathbf{\tilde{x}}_{p}^{0}=\mathbf{x}_{p}^{0}, p\in \mathcal{V},
\\
\nonumber &\mathbf{\tilde{y}}_{i,\alpha}^{0}+\mathbf{\tilde{y}}_{i,\beta}^{0}=\nabla \tilde{f}_i( \mathbf{\tilde{x}}_{i}^{0} ), \,\, \mathbf{\tilde{y}}_{m,\alpha}^{0}+\mathbf{\tilde{y}}_{m,\beta}^{0}=\nabla \tilde{f}_m( \mathbf{\tilde{x}}_{m}^{0} ),
\\
\nonumber &\mathbf{\tilde{y}}_{i,\alpha}^{0}=\mathbf{y}_{i,\alpha}^{0}+\boldsymbol{\delta }_{\alpha}, \,\, \mathbf{\tilde{y}}_{m,\alpha}^{0}=\mathbf{y}_{m,\alpha}^{0}-\boldsymbol{\delta }_{\alpha},
\\
\nonumber &\mathbf{\tilde{y}}_{i,\beta}^{0}=\mathbf{y}_{i,\beta}^{0}+\boldsymbol{\delta }_{\beta}, \,\, \mathbf{\tilde{y}}_{m,\beta}^{0}=\mathbf{y}_{m,\beta}^{0}-\boldsymbol{\delta }_{\beta},
\\
\nonumber &\mathbf{\tilde{y}}_{p,\alpha}^{0}=\mathbf{y}_{p,\alpha}^{0}, \,\, \mathbf{\tilde{y}}_{p,\beta}^{0}=\mathbf{y}_{p,\beta}^{0}, \,\, \forall p\in \mathcal{V}\setminus \{ i,m \},
\end{flalign}
where $\boldsymbol{\delta }_{\alpha}=\left[ \boldsymbol{\delta }_{\alpha}\left( 1 \right) ,\cdots ,\boldsymbol{\delta }_{\alpha}\left( d \right) \right] ^{\top}$ and $\boldsymbol{\delta }_{\beta}=\left[ \boldsymbol{\delta }_{\beta}\left( 1 \right) ,\cdots ,\boldsymbol{\delta }_{\beta}\left( d \right) \right] ^{\top}$ are two arbitrary vectors in $\mathbb{R}^d$, and satisfy $\boldsymbol{\delta }_{\alpha}+\boldsymbol{\delta }_{\beta}=\boldsymbol{\delta }$. Our analysis is divided into two cases: $m\in \mathcal{N}_{i}^{\text{out}}$ and $m\in \mathcal{N}_{i}^{\text{in}}$. Note that if $m\in \mathcal{N}_{i}^{\text{out}}\cap \mathcal{N}_{i}^{\text{in}}$, either of the two cases can be selected to obtain the same result.

\textbf{Case I:} We consider $m\in \mathcal{N}_{i}^{\text{out}}$. One can derive $\mathcal{I}_{\mathcal{A}}( 0:\kappa )=\tilde{\mathcal{I}}_{\mathcal{A}}( 0:\kappa )$ if the parameters are set as follows:
\begin{flalign}
\nonumber \begin{cases}
	\mathbf{\tilde{\Lambda}}_{p}^{0}=\mathbf{\Lambda }_{p}^{0}, \forall p\in \mathcal{V}\setminus \{ i,m \}\\
	\tilde{\gamma}_{i}^{0}( l ) =\gamma _{i}^{0}( l ) \mathbf{y}_{i,\alpha}^{0}( l ) /( \mathbf{y}_{i,\alpha}^{0}( l ) +\boldsymbol{\delta }_{\alpha}( l ) )\\
	\tilde{\gamma}_{m}^{0}( l ) =\gamma _{m}^{0}( l ) \mathbf{y}_{m,\alpha}^{0}( l ) /( \mathbf{y}_{m,\alpha}^{0}( l ) -\boldsymbol{\delta }_{\alpha}( l ) )\\
	\mathbf{\tilde{C}}_{pq}^{0}=\mathbf{C}_{pq}^{0}, \forall p\in \mathcal{V}, q\in \mathcal{V}\setminus \{ i,m \}\\
	\tilde{c}_{pi}^{0}( l ) =c_{pi}^{0}( l ) \mathbf{y}_{i,\alpha}^{0}( l ) /( \mathbf{y}_{i,\alpha}^{0}( l ) +\boldsymbol{\delta }_{\alpha}( l ) ), \forall p\in \mathcal{V}\setminus \{ m \}\\
	\tilde{c}_{mi}^{0}( l ) =( c_{mi}^{0}( l ) \mathbf{y}_{i,\alpha}^{0}( l ) +\boldsymbol{\delta }( l ) ) /( \mathbf{y}_{i,\alpha}^{0}( l ) +\boldsymbol{\delta }_{\alpha}( l ) )\\
	\tilde{c}_{pm}^{0}( l )\!=\!c_{pm}^{0}( l ) \mathbf{y}_{m,\alpha}^{0}( l ) /( \mathbf{y}_{m,\alpha}^{0}( l ) \!-\!\boldsymbol{\delta }_{\alpha}( l ) ), \forall p\in \mathcal{V}\setminus \{ m \}\\
	\tilde{c}_{mm}^{0}( l ) =( c_{mm}^{0}( l ) \mathbf{y}_{m,\alpha}^{0}( l ) -\boldsymbol{\delta }( l ) ) /( \mathbf{y}_{m,\alpha}^{0}( l ) -\boldsymbol{\delta }_{\alpha}( l ) )\\
	\mathbf{\tilde{\Phi}}_{p,\beta}^{0}=\mathbf{\Phi }_{p,\beta}^{0},\forall p\in \mathcal{V}\setminus \{ i,m \}\\
	\tilde{\beta}_{i}^{0}( l ) =( \beta _{i}^{0}( l ) \mathbf{y}_{i,\beta}^{0}( l ) +\boldsymbol{\delta }_{\beta}( l ) ) /( \mathbf{y}_{i,\beta}^{0}( l ) +\boldsymbol{\delta }_{\beta}( l ) )\\
	\tilde{\beta}_{m}^{0}( l ) =( \beta _{m}^{0}( l ) \mathbf{y}_{m,\beta}^{0}( l ) -\boldsymbol{\delta }_{\beta}( l ) ) /( \mathbf{y}_{m,\beta}^{0}( l ) -\boldsymbol{\delta }_{\beta}( l ) )\\
	\mathbf{\tilde{\Phi}}_{p,\alpha}^{0}=\mathbf{\Phi }_{p,\alpha}^{0},\forall p\in \mathcal{V}\setminus \{ i,m \}\\
	\tilde{\alpha}_{i}^{0}( l ) =( \alpha _{i}^{0}( l ) \mathbf{y}_{i,\alpha}^{0}( l ) -\boldsymbol{\delta }_{\beta}( l ) ) /( \mathbf{y}_{i,\alpha}^{0}( l ) +\boldsymbol{\delta }_{\alpha}( l ) )\\
	\tilde{\alpha}_{m}^{0}( l ) =( \alpha _{m}^{0}( l ) \mathbf{y}_{m,\alpha}^{0}( l ) +\boldsymbol{\delta }_{\beta}( l ) ) /( \mathbf{y}_{m,\alpha}^{0}( l ) -\boldsymbol{\delta }_{\alpha}( l ) )\\
	\mathbf{\tilde{R}}_{pq}^{k}=\mathbf{R}_{pq}^{k}, \forall p,q\in \mathcal{V}, k=0,1,\cdots ,\kappa\\
	\mathbf{\tilde{A}}_{pq}^{k}=\mathbf{A}_{pq}^{k}, \forall p,q\in \mathcal{V}, k=0,1,\cdots ,\kappa\\
	\mathbf{\tilde{\Lambda}}_{p}^{k}=\mathbf{\Lambda }_{p}^{k},\forall p\in \mathcal{V}, k=1,2,\cdots ,\kappa\\
	\mathbf{\tilde{C}}_{pq}^{k}=\mathbf{C}_{pq}^{k}, \forall p,q\in \mathcal{V}, k=1,2,\cdots ,\kappa\\
	\mathbf{\tilde{\Phi}}_{p,\beta}^{k}=\mathbf{\Phi }_{p,\beta}^{k},\forall p\in \mathcal{V}, k=1,2,\cdots ,\kappa\\
	\mathbf{\tilde{\Phi}}_{p,\alpha}^{k}=\mathbf{\Phi }_{p,\alpha}^{k},\forall p\in \mathcal{V}, k=1,2,\cdots ,\kappa\\
\end{cases}
\end{flalign}
where $l=1,\cdots ,d$.

\textbf{Case II:} We consider $m\in \mathcal{N}_{i}^{\text{in}}$. One can derive $\mathcal{I}_{\mathcal{A}}( 0:\kappa )=\tilde{\mathcal{I}}_{\mathcal{A}}( 0:\kappa )$ if the parameters are set as follows:
\begin{flalign}
\nonumber \begin{cases}
	\mathbf{\tilde{\Lambda}}_{p}^{0}=\mathbf{\Lambda }_{p}^{0}, \forall p\in \mathcal{V}\setminus \{ i,m \}\\
	\tilde{\gamma}_{i}^{0}( l ) =\gamma _{i}^{0}( l ) \mathbf{y}_{i,\alpha}^{0}( l ) /( \mathbf{y}_{i,\alpha}^{0}( l ) +\boldsymbol{\delta }_{\alpha}( l ) )\\
	\tilde{\gamma}_{m}^{0}( l ) =\gamma _{m}^{0}( l ) \mathbf{y}_{m,\alpha}^{0}( l ) /( \mathbf{y}_{m,\alpha}^{0}( l ) -\boldsymbol{\delta }_{\alpha}( l ) )\\
	\mathbf{\tilde{C}}_{pq}^{0}=\mathbf{C}_{pq}^{0}, \forall p\in \mathcal{V}, q\in \mathcal{V}\setminus \{ i,m \}\\
	\tilde{c}_{pi}^{0}( l ) =c_{pi}^{0}( l ) \mathbf{y}_{i,\alpha}^{0}( l ) /( \mathbf{y}_{i,\alpha}^{0}( l ) +\boldsymbol{\delta }_{\alpha}( l ) ) , \forall p\in \mathcal{V}\setminus \{ i \}\\
	\tilde{c}_{ii}^{0}( l ) =( c_{ii}^{0}( l ) \mathbf{y}_{i,\alpha}^{0}( l ) +\boldsymbol{\delta }( l ) ) /( \mathbf{y}_{i,\alpha}^{0}( l ) +\boldsymbol{\delta }_{\alpha}( l ) )\\
	\tilde{c}_{pm}^{0}( l ) \!=\!c_{pm}^{0}( l ) \mathbf{y}_{m,\alpha}^{0}( l ) /( \mathbf{y}_{m,\alpha}^{0}( l ) \!-\!\boldsymbol{\delta }_{\alpha}( l ) ) ,\forall p\in \mathcal{V}\setminus \{ i \}\\
	\tilde{c}_{im}^{0}( l ) =( c_{im}^{0}( l ) \mathbf{y}_{m,\alpha}^{0}( l ) -\boldsymbol{\delta }( l ) ) /( \mathbf{y}_{m,\alpha}^{0}( l ) -\boldsymbol{\delta }_{\alpha}( l ) )\\
	\mathbf{\tilde{\Phi}}_{p,\beta}^{0}=\mathbf{\Phi }_{p,\beta}^{0},\forall p\in \mathcal{V}\setminus \{ i,m \}\\
	\tilde{\beta}_{i}^{0}( l ) =( \beta _{i}^{0}( l ) \mathbf{y}_{i,\beta}^{0}( l ) +\boldsymbol{\delta }_{\beta}( l ) ) /( \mathbf{y}_{i,\beta}^{0}( l ) +\boldsymbol{\delta }_{\beta}( l ) )\\
	\tilde{\beta}_{m}^{0}( l ) =( \beta _{m}^{0}( l ) \mathbf{y}_{m,\beta}^{0}( l ) -\boldsymbol{\delta }_{\beta}( l ) ) /( \mathbf{y}_{m,\beta}^{0}( l ) -\boldsymbol{\delta }_{\beta}( l ) )\\
	\mathbf{\tilde{\Phi}}_{p,\alpha}^{0}=\mathbf{\Phi }_{p,\alpha}^{0},\forall p\in \mathcal{V}\setminus \{ i,m \}\\
	\tilde{\alpha}_{i}^{0}( l ) =( \alpha _{i}^{0}( l ) \mathbf{y}_{i,\alpha}^{0}( l ) -\boldsymbol{\delta }_{\beta}( l ) ) /( \mathbf{y}_{i,\alpha}^{0}( l ) +\boldsymbol{\delta }_{\alpha}( l ) )\\
	\tilde{\alpha}_{m}^{0}( l ) =( \alpha _{m}^{0}( l ) \mathbf{y}_{m,\alpha}^{0}( l ) +\boldsymbol{\delta }_{\beta}( l ) ) /( \mathbf{y}_{m,\alpha}^{0}( l ) -\boldsymbol{\delta }_{\alpha}( l ) )\\
	\mathbf{\tilde{R}}_{pq}^{k}=\mathbf{R}_{pq}^{k}, \forall p,q\in \mathcal{V}, k=0,1,\cdots ,\kappa\\
	\mathbf{\tilde{A}}_{pq}^{k}=\mathbf{A}_{pq}^{k}, \forall p,q\in \mathcal{V}, k=0,1,\cdots ,\kappa\\
	\mathbf{\tilde{\Lambda}}_{p}^{k}=\mathbf{\Lambda }_{p}^{k},\forall p\in \mathcal{V}, k=1,2,\cdots ,\kappa\\
	\mathbf{\tilde{C}}_{pq}^{k}=\mathbf{C}_{pq}^{k}, \forall p,q\in \mathcal{V}, k=1,2,\cdots ,\kappa\\
	\mathbf{\tilde{\Phi}}_{p,\beta}^{k}=\mathbf{\Phi }_{p,\beta}^{k},\forall p\in \mathcal{V}, k=1,2,\cdots ,\kappa\\
	\mathbf{\tilde{\Phi}}_{p,\alpha}^{k}=\mathbf{\Phi }_{p,\alpha}^{k},\forall p\in \mathcal{V}, k=1,2,\cdots ,\kappa\\
\end{cases}
\end{flalign}
where $l=1,\cdots ,d$.

Next we explain why the parameter settings in Case I guarantee that $\mathcal{I}_{\mathcal{A}}( 0:\kappa )=\tilde{\mathcal{I}}_{\mathcal{A}}( 0:\kappa )$. We consider $k=0$ and $1 \le k \le \kappa$, separately. At $k=0$, one can verify that the parameter settings do not violate the requirements of TABLE I, and the accessible information to corrupted agents $\mathcal{A}$ remains unchanged. Moreover, under these settings, it holds $\mathbf{\tilde{x}}_{p}^{1}=\mathbf{x}_{p}^{1}$, $\mathbf{\tilde{y}}_{p,\alpha}^{1}=\mathbf{y}_{p,\alpha}^{1}$, and $\mathbf{\tilde{y}}_{p,\beta}^{1}=\mathbf{y}_{p,\beta}^{1}$ for every $p\in \mathcal{V}$. Then, for $1 \le k \le \kappa$, the parameter settings are the same under $\nabla \tilde{f}_i( \cdot )$ and $\nabla f_i( \cdot )$, and thus the accessible information to corrupted agents $\mathcal{A}$ also remains unchanged. In line with a similar discussion, the parameter settings in Case II also guarantee that $\mathcal{I}_{\mathcal{A}}( 0:\kappa )=\tilde{\mathcal{I}}_{\mathcal{A}}( 0:\kappa )$.

To summarize, since $\mathcal{I}_{\mathcal{A}}( 0:\kappa )=\tilde{\mathcal{I}}_{\mathcal{A}}( 0:\kappa )$, we have $\nabla f_i( \cdot ), \nabla \tilde{f}_i( \cdot ) \in \nabla _i( \mathcal{I}_{\mathcal{A}}( 0:\kappa ) )$. Thus, for any $\kappa \in \mathbb{N}$,
\begin{flalign}
\nonumber \mathcal{D}_i( \mathcal{I}_{\mathcal{A}}( 0:\kappa ) ) \ge \underset{\boldsymbol{\delta }\in \mathbb{R}^d}{\rm{sup}}\lVert \nabla f_i( \cdot ) -( \nabla f_i( \cdot ) +\boldsymbol{\delta } ) \rVert =\infty.
\end{flalign}
Therefore, PPSD can preserve the privacy of agent $i$ if $\mathcal{N}_{i}^{\text{out}}\cup \mathcal{N}_{i}^{\text{in}}\not\subset \mathcal{A}$.
\end{proof}
\end{theorem}

\begin{remark}
Note that there are infinite solutions to set the parameter to ensure that $\mathcal{I}_{\mathcal{A}}\left( 0:\kappa \right)=\tilde{\mathcal{I}}_{\mathcal{A}}\left( 0:\kappa \right)$. The discussion above just offers one such solution.
\end{remark}

Note that if all neighbors of normal agent $i$ are corrupt agents, i.e., $\mathcal{N}_{i}^{\text{out}}\cup \mathcal{N}_{i}^{\text{in}}\subset \mathcal{A}$, the private information of agent $i$ can be duduced by the corrupted agents.
\begin{theorem}
Let the parameters satisfy TABLE I. In PPSD, the gradient information of each normal agent $i$ can be inferred by corrupted agents $\mathcal{A}$ if $\mathcal{N}_{i}^{\text{out}}\cup \mathcal{N}_{i}^{\text{in}}\subset \mathcal{A}$.
\begin{proof}
From \eqref{Eq:3}, one can derive
\begin{flalign}
\label{Eq:15} \mathbf{y}_{i,\alpha}^{k+1}=\mathbf{C}_{ii}^{k}\mathbf{y}_{i,\alpha}^{k}+\sum_{j\in \mathcal{N}_{i}^{\text{in}}}{\mathbf{C}_{ij}^{k}\mathbf{y}_{j,\alpha}^{k}}+( \mathbf{I}_d-\mathbf{\Phi }_{i,\beta}^{k} ) \mathbf{y}_{i,\beta}^{k}. \tag{15}
\end{flalign}
Then, recalling the settings of $\mathbf{C}_{ji}^{k}$ and $\mathbf{\Phi }_{i,\alpha}^{k}$ in TABLE I, we have
\begin{flalign}
\label{Eq:16} \mathbf{y}_{i,\alpha}^{k}=\mathbf{C}_{ii}^{k}\mathbf{y}_{i,\alpha}^{k}+\mathbf{\Phi }_{i,\alpha}^{k}\mathbf{y}_{i,\alpha}^{k}+\sum_{j\in \mathcal{N}_{i}^{\text{out}}}{\mathbf{C}_{ji}^{k}\mathbf{y}_{i,\alpha}^{k}}.\tag{16}
\end{flalign}
Moreover, using the relations \eqref{Eq:3}, \eqref{Eq:4}, \eqref{Eq:15}, and \eqref{Eq:16} yields
\begin{flalign}
\nonumber \mathbf{y}_{i,\alpha}^{k+1}+\mathbf{y}_{i,\beta}^{k+1}=&\mathbf{C}_{ii}^{k}\mathbf{y}_{i,\alpha}^{k}+\sum_{j\in \mathcal{N}_{i}^{\text{in}}}{\mathbf{C}_{ij}^{k}\mathbf{y}_{j,\alpha}^{k}}+\mathbf{y}_{i,\beta}^{k}
\\
\label{Eq:17}\,\,                &+\mathbf{\Phi }_{i,\alpha}^{k}\mathbf{y}_{i,\alpha}^{k}+\nabla f_i( \mathbf{x}_{i}^{k+1} ) -\nabla f_i( \mathbf{x}_{i}^{k} ), \tag{17}
\end{flalign}
\begin{flalign}
\label{Eq:18} \mathbf{y}_{i,\alpha}^{k}\!+\!\mathbf{y}_{i,\beta}^{k}\!=\!\mathbf{C}_{ii}^{k}\mathbf{y}_{i,\alpha}^{k}\!+\!\mathbf{\Phi }_{i,\alpha}^{k}\mathbf{y}_{i,\alpha}^{k}\!+\!\sum_{j\in \mathcal{N}_{i}^{\text{out}}}{\mathbf{C}_{ji}^{k}\mathbf{y}_{i,\alpha}^{k}}\!+\!\mathbf{y}_{i,\beta}^{k}. \tag{18}
\end{flalign}
Define $\mathbf{z}_{i}^{k}\coloneqq \mathbf{y}_{i,\alpha}^{k}+\mathbf{y}_{i,\beta}^{k}$. One can verify $\mathbf{z}_{i}^{0}=\nabla f_i( \mathbf{x}_{i}^{0} )$ for any $i\in \mathcal{V}$. Subtracting \eqref{Eq:17} from \eqref{Eq:18} gives
\begin{flalign}
\nonumber \mathbf{z}_{i}^{k+1}\!-\!\mathbf{z}_{i}^{k}\!=\!\sum_{j\in \mathcal{N}_{i}^{\text{in}}}{\mathbf{C}_{ij}^{k}\mathbf{y}_{j,\alpha}^{k}}\!-&\!\sum_{j\in \mathcal{N}_{i}^{\text{out}}}{\mathbf{C}_{ji}^{k}\mathbf{y}_{i,\alpha}^{k}}
\\
\label{Eq:19} &+\nabla f_i( \mathbf{x}_{i}^{k+1} ) -\nabla f_i( \mathbf{x}_{i}^{k} ). \tag{19}
\end{flalign}
Computing \eqref{Eq:19} recursively, we have
\begin{flalign}
\label{Eq:20} \mathbf{z}_{i}^{k+\!1}\!=\!\sum_{t=0}^k{\!\!\Big( \sum_{j\in \mathcal{N}_{i}^{\text{in}}}\!{\!\mathbf{C}_{ij}^{t}\mathbf{y}_{j,\alpha}^{t}}\!\!-\!\!\sum_{j\in \mathcal{N}_{i}^{\text{out}}}\!{\!\mathbf{C}_{ji}^{t}\mathbf{y}_{i,\alpha}^{t}} \!\Big)}\!+\!\nabla\! f_i( \mathbf{x}_{i}^{k+\!1} ), \tag{20}
\end{flalign}
where the first term on the right side of \eqref{Eq:20} is accessible to all corrupted agents.

Define $\mathbf{z}^k=[ ( \mathbf{z}_{1}^{k} ) ^{\top},\cdots ,( \mathbf{z}_{n}^{k} ) ^{\top} ] ^{\top}$. One can verify that $( \mathbf{1}_{n}^{\top}\otimes \mathbf{I}_d ) \mathbf{z}^k=( \mathbf{1}_{2n}^{\top}\otimes \mathbf{I}_d ) \mathbf{y}^k$ holds. Since $\lim _{k\rightarrow \infty}\mathbf{y}_{i}^{k+1}=\mathbf{0}_d$, it holds $\lim _{k\rightarrow \infty}\mathbf{z}_{i}^{k+1}=\mathbf{0}_d$. Thus, we have
\begin{flalign}
\nonumber \lim _{k\rightarrow \infty}\!\nabla\!f_i( \mathbf{x}_{i}^{k+1} ) \!=\!-\!\lim _{k\rightarrow \infty}\!\sum_{t=0}^k\!{\Big(\! \sum_{j\in \mathcal{N}_{i}^{\text{in}}}\!{\!\mathbf{C}_{ij}^{t}\mathbf{y}_{j,\alpha}^{t}}\!-\!\sum_{j\in \mathcal{N}_{i}^{\text{out}}}\!{\!\mathbf{C}_{ji}^{t}\mathbf{y}_{i,\alpha}^{t}}\! \Big)}.
\end{flalign}
Since $\lim _{k\rightarrow \infty}\mathbf{x}_{i}^{k+1}=\lim _{k\rightarrow \infty}\mathbf{x}_{j}^{k+1}=\mathbf{x}^{\ast}$, each corrupted agent $j$ also grasps $\lim _{k\rightarrow \infty}\mathbf{x}_{i}^{k+1}$, implying that all corrupted agents are capable of inferring the gradient information of agent $i$ at $\mathbf{x}^{\ast}$. Consequently, the gradient information of each normal agent $i$ can be inferred by corrupted agents $\mathcal{A}$ when $\mathcal{N}_{i}^{\text{out}}\cup \mathcal{N}_{i}^{\text{in}}\subset \mathcal{A}$.
\end{proof}
\end{theorem}

\begin{remark}
By Theorem 1, we can know that the parameter settings in TABLE I have no impact on  the convergence accuracy. Moreover, from the proof of Theorem 2, one can see that simply designing the values of the related parameters at $k=0$ is enough to mask the gradient information.
\end{remark}

\textbf{Discussion:} Our work can be extended to preserve private information under eavesdropping attacks, where an external eavesdropper exists, which can capture all sharing information by wiretapping communication channels.

By the proof of Theorem 2, it is known that arbitrary variations of agent $i$'s gradient can be completely compensated by changing $\{ \mathbf{C}_{mi}^{0},\mathbf{\Phi }_{i,\alpha}^{0},\mathbf{\Phi }_{i,\beta}^{0},\mathbf{\Lambda }_{i}^{0},\mathbf{\Lambda }_{m}^{0},\mathbf{C}_{mm}^{0} \}$ or $\{ \mathbf{C}_{im}^{0},\mathbf{\Phi }_{i,\alpha}^{0},\mathbf{\Phi }_{i,\beta}^{0},\mathbf{\Lambda }_{i}^{0},\mathbf{\Lambda }_{m}^{0},\mathbf{C}_{ii}^{0} \}$ that are invisible to the eavesdropper.

\begin{assumption}
For the network $\mathcal{G}=( \mathcal{V},\mathcal{E} )$, each agent $i \in \mathcal{V}$ has at least one neighbor $m\in \mathcal{N}_{i}^{\text{out}}\cup \mathcal{N}_{i}^{\text{in}}$ whose interactive information $\mathbf{C}_{mi}^{0}\mathbf{y}_{m}^{0}$ or $\mathbf{C}_{im}^{0}\mathbf{y}_{m}^{0}$ with $i$ is inaccessible to the eavesdropper.
\end{assumption}

\begin{theorem}
Let the parameters satisfy TABLE I and Assumption 3 hold. In PPSD, the gradient information of each agent $i\in \mathcal{V}$ can be preserved against the external eavesdropper.
\begin{proof}
Pursuing a similar path of proof in Theorem 2, one can know that there always exists feasible parameters such that the information accessed by an eavesdropper under $\nabla \tilde{f}_i( \cdot ) =\nabla f_i( \cdot ) +\boldsymbol{\delta }$ (where $\boldsymbol{\delta }$ is an arbitrary vector in $\mathbb{R}^d$) is the same as the one under $\nabla f_i( \cdot )$. Thus, the eavesdropper has no way of inferring which is the true gradient from the infinite arbitrary variants of $\nabla f_i( \cdot )$ based on the available information.
\end{proof}
\end{theorem}

\section{Simulation Verification}
In this section, we test the privacy and convergence performances of PPSD in two practical problems. Here, we constructed two networks with 5 and 500 nodes, respectively, for simulations, see Fig. 2.

\begin{figure}[htbp]
\centering
\subfloat[$\mathcal{G}_1$]{
\label{fig:2.a}
\includegraphics[width=4.3cm]{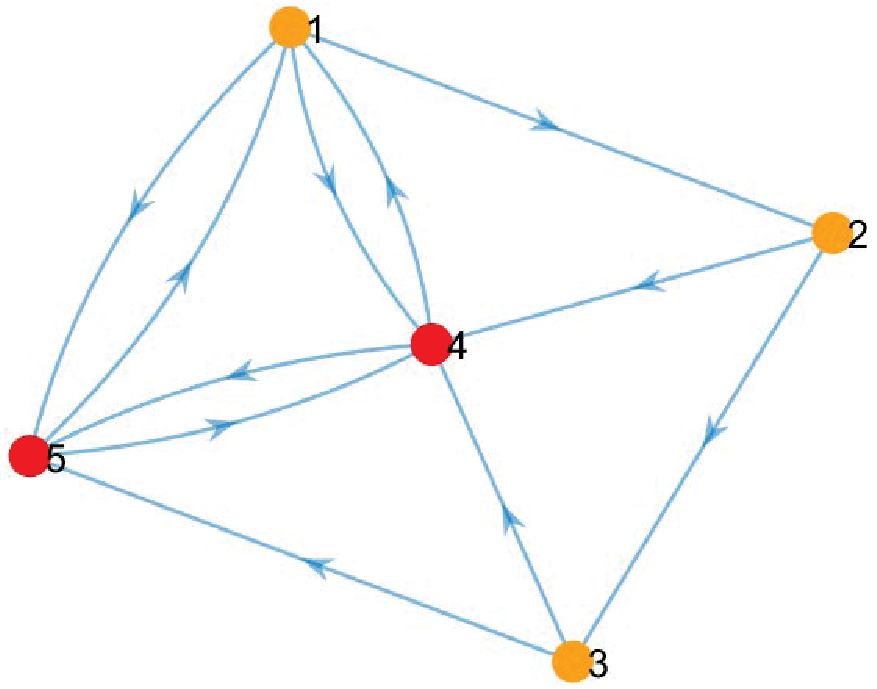}}
\subfloat[$\mathcal{G}_2$]{
\label{fig:2.b}
\includegraphics[width=4.3cm]{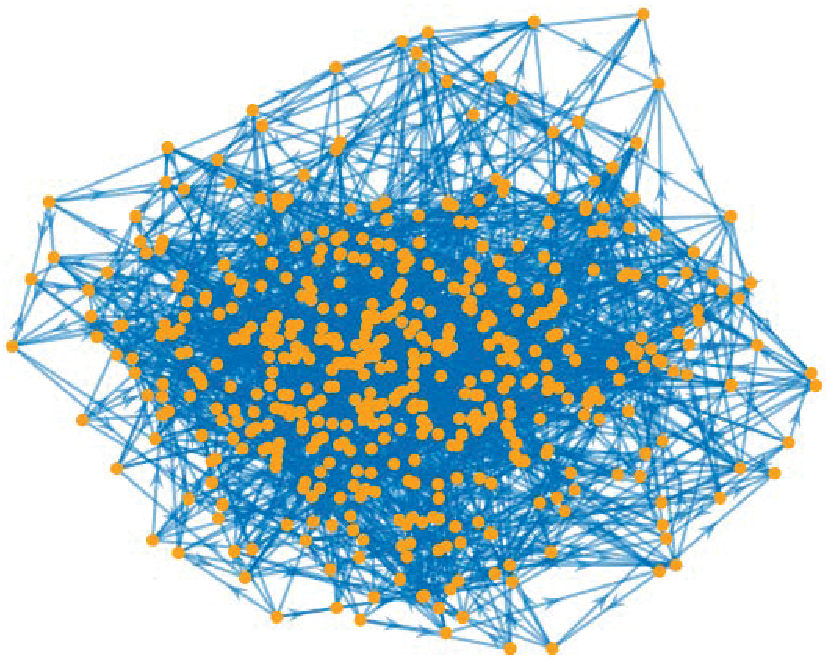}}
\caption{Networks with $5$ agents (a) and $500$ agents (b).}
\label{fig:2}
\end{figure}

\subsection{Privacy Preservation in the Rendezvous Problem}
In this problem, the task of each agent $i$ is to collaborate with other agents to find the nearest rendezvous point without revealing its initial position $\mathbf{p}_i\in \mathbb{R}^d$. Each $f_i$ of the rendezvous problem is modeled as $f_i( \mathbf{x} ) =\frac{1}{2}\lVert \mathbf{x}-\mathbf{p}_i \rVert ^2$.

We set $d=1$ and choose the network $\mathcal{G}_1$ as shown in Fig. 2(a). Let agents $4$ and $5$ be corrupted agents, i.e., $\mathcal{A}=\{ 4,5 \}$, which try to infer the gradient information of normal agent $1$, and agent $2$ be a normal out-neighbor of agent $1$. Note that agent $4$ and agent $5$ can collude with each other. In this experiment, we first run PPSD once and record the accessible information to the corrupted agents $\mathcal{I}_{\mathcal{A}}=\{ \mathcal{I}_4( k ) \cup \mathcal{I}_5( k ) \left| k=0,1\cdots \right. \}$. Then, we show that the information accessed by the corrupted agents, denoted by $\tilde{\mathcal{I}}_{\mathcal{A}}$, under the gradient $\nabla \tilde{f}_1( \mathbf{\tilde{x}}_1 ) =\nabla f_1( \mathbf{x}_1 ) +\boldsymbol{\delta }$, where each element of $\boldsymbol{\delta }$ is selected from $(0, 5000)$ arbitrarily.

Fig. 3 shows that the gradients $\nabla f_1( \mathbf{x}_1 )$ and $\nabla \tilde{f}_1( \mathbf{\tilde{x}}_1 )$ used in the two runs of PPSD are clearly different, but the information accessible to the corrupted agents $\mathcal{A}$ (i.e., $\mathbf{x}_{1}^{k}$, $\mathbf{\Lambda }_{1}^{k}\mathbf{y}_{1,\alpha}^{k}$, $\mathbf{C}_{41}^{k}\mathbf{y}_{1,\alpha}^{k}$, $\mathbf{C}_{51}^{k}\mathbf{y}_{1,\alpha}^{k}$ in $\mathcal{I}_{\mathcal{A}}$ and $\mathbf{\tilde{x}}_{1}^{k}$, $\mathbf{\tilde{\Lambda}}_{1}^{k}\mathbf{\tilde{y}}_{1,\alpha}^{k}$, $\mathbf{\tilde{C}}_{41}^{k}\mathbf{\tilde{y}}_{1,\alpha}^{k}$, $\mathbf{\tilde{C}}_{51}^{k}\mathbf{\tilde{y}}_{1,\alpha}^{k}$ in $\tilde{\mathcal{I}}_{\mathcal{A}}$) is exactly the same as shown in Fig. 4. Therefore, the corrupted agents $\mathcal{A}$ are unable to infer which is the true gradient information of normal agent $1$.

\begin{figure}[htbp]
\centering
\includegraphics[width=3.1in]{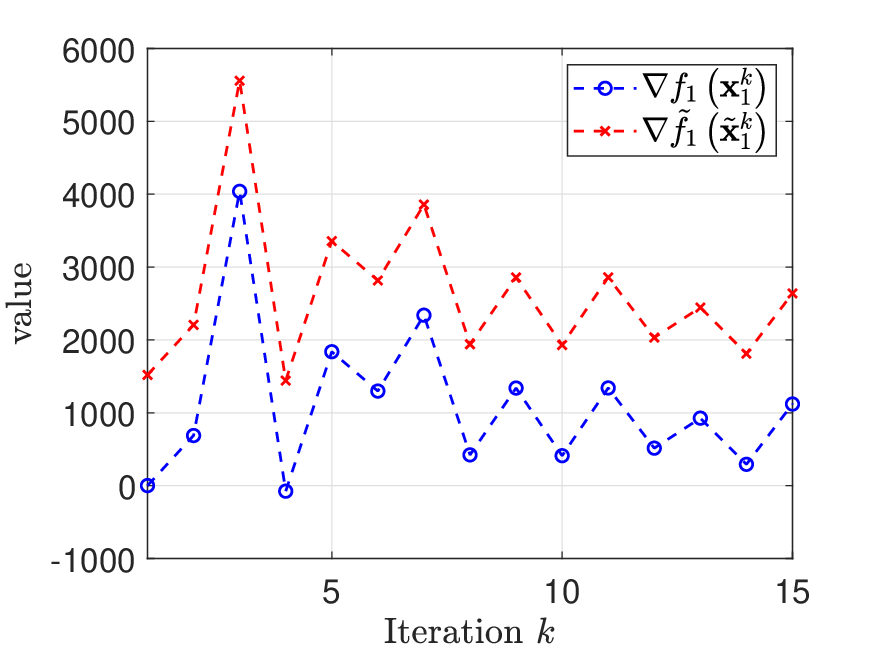}
\caption{Two different gradients of agent $1$.}
\label{fig:3}
\end{figure}

\begin{figure}[htbp]
\begin{minipage}[t]{0.5\textwidth}
\centering
\label{fig:4.a}
\includegraphics[width=3.1in]{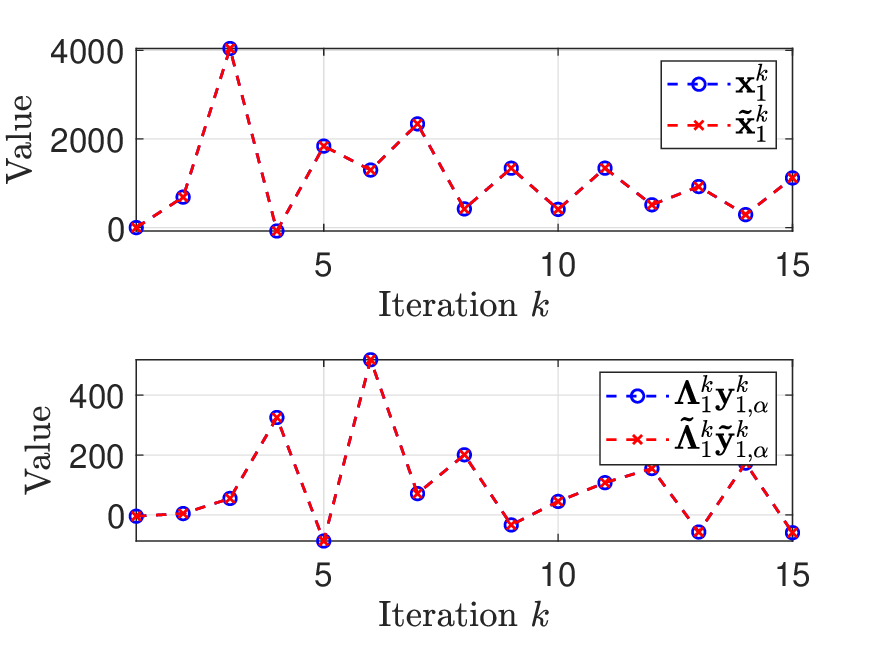}
\end{minipage}
\quad
\begin{minipage}[t]{0.5\textwidth}
\centering
\label{fig:4.b}
\includegraphics[width=3.1in]{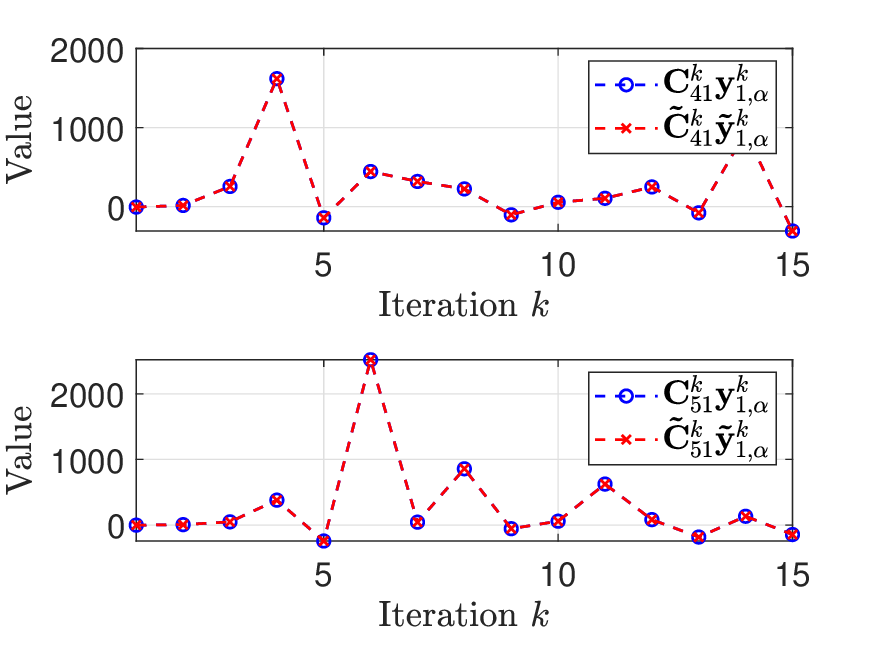}
\end{minipage}
\caption{The information accessible to corrupted agents $4$ and $5$ are the same under two different gradients of agent $1$ shown in Fig. 3.}
\label{fig:4}
\end{figure}

\subsection{Decentralized Linear Regression}
We show the convergence performances of PPSD in the decentralized linear regression problem, where each agent collaborates with each other to estimate an unknown signal $\mathbf{s}_0\in \mathbb{R}^d$. Specifically, each agent grasps a measurement relation $\mathbf{m}_i=\boldsymbol{Q}_i\mathbf{s}_0+\boldsymbol{\zeta }_i$, in which $\boldsymbol{Q}_i\in \mathbb{R}^{p_i\times d}$ is an observation matrix and $\boldsymbol{\zeta }_i \in \mathbb{R}^{p_i}$ is an interfering noise. Each $f_i$ of the linear regression problem is modeled as $f_i( \mathbf{x} ) =\lVert \boldsymbol{Q}_i\mathbf{x}-\mathbf{m}_i \rVert ^2$.

We set $d=10$ and $p_i=10$ for all $i\in \mathcal{V}$, and choose the network $\mathcal{G}_1$. Let each element of $\mathbf{s}_0$ and $\boldsymbol{\zeta }_i$ be i.i.d. random value drawn from $\mathcal{N}( 0,1 )$ and $\mathcal{N}( 0,0.2 )$, respectively. We fix $\boldsymbol{Q}_i$ with each element being i.i.d. random value drawn from $\mathcal{N}( 0,1 )$ and then normalize the matrix. The residual $\lVert \mathbf{x}^k-\mathbf{1}_n\otimes \mathbf{x}^{\ast} \rVert$ is used as the performance metric.

\textbf{Comparison with Decentralized Optimization Algorithms.} We compare PPSD with Push-Pull \cite{Pu2021}, $\mathcal{A}\mathcal{B}$ \cite{Xin2018b}, Push-DIGing \cite{Nedic2017}, ADD-OPT \cite{Xi2017b}, and Subgradient-Push \cite{Nedic2015} to verify the impact of security mechanisms on convergence accuracy. As shown in Fig. 5, PPSD achieves a linear convergence rate and the place marked in the curve indicates that the randomness added to the corresponding parameters prolongs the convergence process.

\begin{figure}[htbp]
\centering
\includegraphics[width=3.1in]{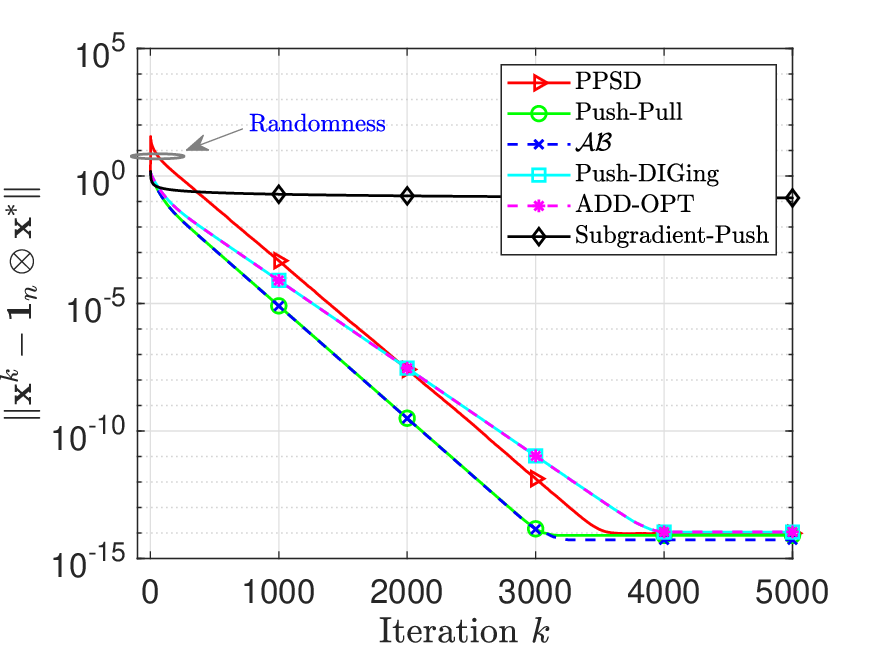}
\caption{Performance comparison.}
\label{fig:5}
\end{figure}

\textbf{Comparison with \cite{Chen2023a}.} We compare PPSD with the differential privacy based algorithm \cite{Chen2023a}. We consider the performance of the algorithm \cite{Chen2023a} under four privacy levels, i.e., $\sigma =10^{-6},10^{-4},10^{-2},1$, where $\sigma$ is the scale parameter. Note that a larger $\sigma$ means a greater privacy level. From Fig. 6, one can see that the differential privacy based algorithm \cite{Chen2023a} has a compromise between convergence accuracy and privacy level, and also demonstrates the advantages of PPSD in ensuring convergence accuracy.

\begin{figure}[htbp]
\centering
\includegraphics[width=3.1in]{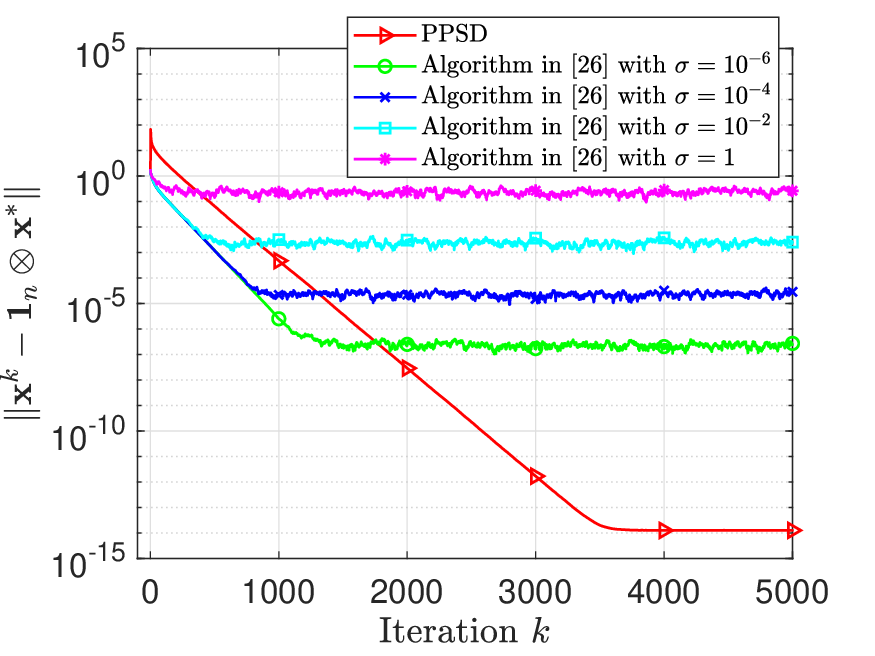}
\caption{Performance comparison.}
\label{fig:6}
\end{figure}

\textbf{Comparison with \cite{Gao2023}.} We compare PPSD with the dynamics based privacy-preserving algorithm \cite{Gao2023}. The main idea of the algorithm in \cite{Gao2023} to achieve privacy preservation is to add randomness to the mixing matrix in the first $K$ iterations. Here, we consider four cases: $K=1,2,3,4$. As depicted in Fig. 7, the randomness added in the first $K$ iterations in \cite{Gao2023} has no impact on the convergence rate, but there is a significant decay in the convergence accuracy. Although PPSD converges slightly slower than the algorithm \cite{Gao2023}, it ensures a higher convergence accuracy.

\begin{figure}[htbp]
\centering
\includegraphics[width=3.1in]{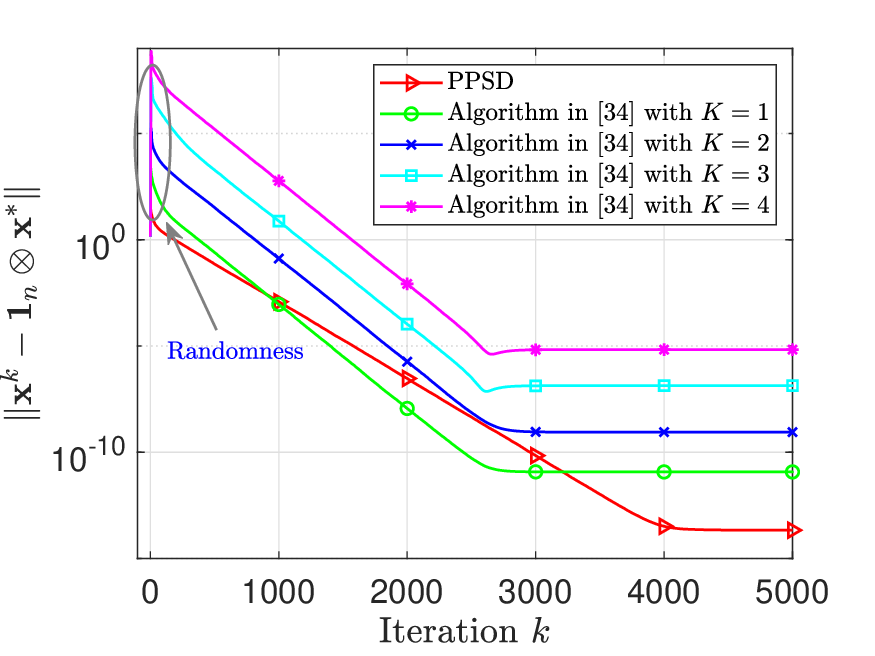}
\caption{Performance comparison.}
\label{fig:7}
\end{figure}

\textbf{Simulation on the Large-scale Network.} Finally, we use $\mathcal{G}_2$ to check the scalability of PPSD. Other settings are the same as the above. It is shown in Fig. 8 that PPSD achieves a linear convergence rate even in large-scale networks.
\begin{figure}[htbp]
\centering
\includegraphics[width=3.1in]{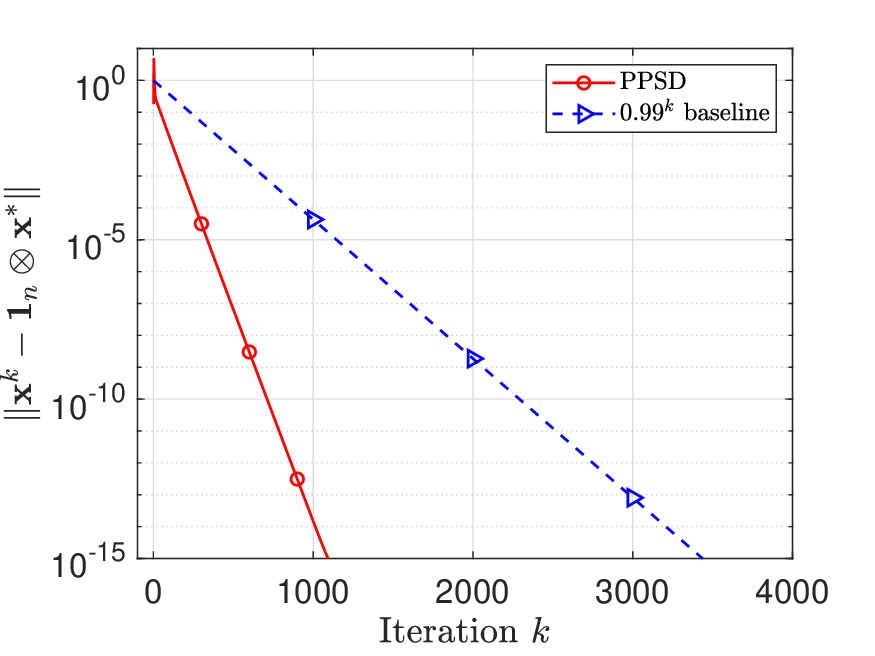}
\caption{The performance on $\mathcal{G}_2$.}
\label{fig:8}
\end{figure}

\section{Conclusion}
We proposed a novel privacy-preserving decentralized optimization algorithm (PPSD) via state decomposition for unbalanced digraphs. Compared to algorithms based on differential privacy or homomorphic encryption, PPSD ensures convergence performance without incurring additional computational burden. We critically analyzed the convergence rate and privacy performance of PPSD and further verified it in simulation experiments. Future work will continually focus on the design of privacy mechanisms for decentralized algorithms.


\newpage
\clearpage
\appendices

\section{Proof of Lemma 4}
\begin{proof}
By the dynamics \eqref{Eq:10}, it follows for $k \ge \bar{N}$:
\begin{flalign}
\nonumber \mathbf{x}^{k+1}=&\mathbf{R}_{k:k-\bar{N}+1}\mathbf{x}^{k-\bar{N}+1}
\\
\nonumber &-\gamma \Big( \mathbf{A}^k\mathbf{Ty}^k+\sum_{l=1}^{\bar{N}-1}{\mathbf{R}_{k:k-l+1}\mathbf{A}^{k-l}\mathbf{Ty}^{k-l}} \Big)
\end{flalign}
Then, for $\mathbf{\tilde{x}}_{\text{w}}^{k+1}$, it holds
\begin{flalign}
\nonumber &\lVert \mathbf{\tilde{x}}_{\text{w}}^{k+1} \rVert
\\
\nonumber \le &\lVert [ ( \mathbf{I}_n-\mathbf{1}_n( \boldsymbol{\phi }^{k+1} ) ^{\top} ) \otimes \mathbf{I}_d ] \mathbf{R}_{k:k-\bar{N}+1}\mathbf{x}^{k-\bar{N}+1} \rVert
\\
\nonumber \,\,           &+\gamma \lVert [ ( \mathbf{I}_n-\mathbf{1}_n( \boldsymbol{\phi }^{k+1} ) ^{\top} ) \otimes \mathbf{I}_d ] \mathbf{A}^k\mathbf{Ty}^k \rVert
\\
\nonumber \,\,           &+\!\gamma \sum_{l=1}^{\bar{N}-1}{\lVert [ ( \mathbf{I}_n\!-\!\mathbf{1}_n( \boldsymbol{\phi }^{k+1} ) ^{\top} )\! \otimes\! \mathbf{I}_d ] \mathbf{R}_{k:k-l+1}\mathbf{A}^{k-l}\mathbf{Ty}^{k-l} \rVert}
\\
\nonumber \,\,        \le &r_R\lVert [ ( \mathbf{I}_n\!-\!\mathbf{1}_n( \boldsymbol{\phi }^{k-\bar{N}+1} ) ^{\top} )\! \otimes\! \mathbf{I}_d ] \mathbf{x}^{k-\bar{N}+1} \rVert \!+\!\gamma Q_R\sqrt{n}\lVert \mathbf{y}^k \rVert
\\
\nonumber \,\,           &+\gamma \sum_{l=1}^{\bar{N}-1}{\lVert [ ( \mathbf{I}_n-\mathbf{1}_n( \boldsymbol{\phi }^{k-l+1} ) ^{\top} ) \otimes \mathbf{I}_d ] \mathbf{A}^{k-l}\mathbf{Ty}^{k-l} \rVert}
\\
\label{Eq:A2}\,\,        \le &r_R\lVert \mathbf{\tilde{x}}_{\text{w}}^{k-\bar{N}+1} \rVert +\gamma Q_R\sqrt{n}\sum_{l=0}^{\bar{N}-1}{\lVert \mathbf{y}^{k-l} \rVert}, \tag{A2}
\end{flalign}
where the second inequality uses that $\lVert \mathbf{I}_n-\mathbf{1}_n( \boldsymbol{\phi }^{k-\bar{N}+1} ) ^{\top} \rVert \le 2\sqrt{n}$, $\lVert \mathbf{A}^k \rVert \le \sqrt{n}$, $2\sqrt{n}\le Q_R$, $\lVert \mathbf{T} \rVert =1$, and $\bar{N}=\max \{ N_R,N_P \}$. Next, we analyze $\lVert \mathbf{y}^k \rVert$. From the dynamic of \eqref{Eq:11}, we have
\begin{flalign}
\nonumber\,\,  &( \mathbf{1}_{\tilde{n}}( \mathbf{v}^k ) ^{\top}\otimes \mathbf{I}_d ) \mathbf{s}^k
\\
\nonumber =&( \mathbf{1}_{\tilde{n}}( \mathbf{v}^{k-1} ) ^{\top}\!\otimes\! \mathbf{I}_d ) \mathbf{s}^{k-1}\!+\!( \mathbf{1}_{\tilde{n}}\mathbf{1}_{\tilde{n}}^{\top}\!\otimes\! \mathbf{I}_d ) ( \nabla \mathbf{\hat{f}}( \mathbf{x}^k ) \!-\!\nabla \mathbf{\hat{f}}( \mathbf{x}^{k-1} ) )
\\
\nonumber =&( \mathbf{1}_{\tilde{n}}( \mathbf{v}^1 ) ^{\top}\otimes \mathbf{I}_d ) \mathbf{s}^1+( \mathbf{1}_{\tilde{n}}\mathbf{1}_{\tilde{n}}^{\top}\otimes \mathbf{I}_d ) ( \nabla \mathbf{\hat{f}}( \mathbf{x}^k ) -\nabla \mathbf{\hat{f}}( \mathbf{x}^1 ) )
\\
\nonumber =&( \mathbf{1}_{\tilde{n}}\mathbf{1}_{\tilde{n}}^{\top}\otimes \mathbf{I}_d ) ( \mathbf{y}^1-\nabla \mathbf{\hat{f}}( \mathbf{x}^1 ) ) +( \mathbf{1}_{\tilde{n}}\mathbf{1}_{\tilde{n}}^{\top}\otimes \mathbf{I}_d ) \nabla \mathbf{\hat{f}}( \mathbf{x}^k )
\\
\nonumber =&( \mathbf{1}_{\tilde{n}}\mathbf{1}_{\tilde{n}}^{\top}\otimes \mathbf{I}_d ) \nabla \mathbf{\hat{f}}( \mathbf{x}^k ) =( \mathbf{1}_n\mathbf{1}_{n}^{\top}\otimes \mathbf{I}_d ) \nabla \mathbf{f}( \mathbf{x}^k ),
\end{flalign}
where the last third relation holds since $( \mathbf{1}_{\tilde{n}}( \mathbf{v}^1 ) ^{\top}\otimes \mathbf{I}_d ) \mathbf{s}^1=( \mathbf{1}_{\tilde{n}}( \mathbf{v}^1 ) ^{\top}( \mathbf{\bar{V}}^1 ) ^{-1}\otimes \mathbf{I}_d ) \mathbf{y}^1=( \mathbf{1}_{\tilde{n}}\mathbf{1}_{\tilde{n}}^{\top}\otimes \mathbf{I}_d ) \mathbf{y}^1$. From the definition of $\mathbf{\tilde{s}}_{\mathbf{w}}^{k}$ and the optimality condition $\sum\nolimits_{i=1}^n{\nabla f_i( \mathbf{x}^{\ast} )}=\mathbf{0}_d$, we obtain
\begin{flalign}
\nonumber \mathbf{s}^k=&\mathbf{\tilde{s}}_{\text{w}}^{k}+( \mathbf{1}_{\tilde{n}}( \mathbf{v}^k ) ^{\top}\otimes \mathbf{I}_d ) \mathbf{s}^k
\\
\nonumber\,\, =&\mathbf{\tilde{s}}_{\text{w}}^{k}+( \mathbf{1}_n\mathbf{1}_{n}^{\top}\otimes \mathbf{I}_d ) \nabla \mathbf{f}( \mathbf{x}^k )
\\
\nonumber\,\, =&\mathbf{\tilde{s}}_{\text{w}}^{k}+( \mathbf{1}_n\mathbf{1}_{n}^{\top}\otimes \mathbf{I}_d ) ( \nabla \mathbf{f}( \mathbf{x}^k ) -\nabla \mathbf{f}( \mathbf{1}_n\otimes \mathbf{x}^{\ast} ) ).
\end{flalign}
Thus, we bound $\mathbf{s}^k$ as
\begin{flalign}
\nonumber \lVert \mathbf{s}^k \rVert \le& \lVert \mathbf{\tilde{s}}_{\text{w}}^{k} \rVert +nL\lVert \mathbf{x}^k-\mathbf{1}_n\otimes \mathbf{x}^{\ast} \rVert
\\
\nonumber      \le& \lVert \mathbf{\tilde{s}}_{\text{w}}^{k} \rVert +nL\lVert \mathbf{x}^k-( \mathbf{1}_n( \boldsymbol{\phi }^k ) ^{\top}\otimes \mathbf{I}_d ) \mathbf{x}_k \rVert
\\
\nonumber         &+nL\lVert ( \mathbf{1}_n( \boldsymbol{\phi }^k ) ^{\top}\otimes \mathbf{I}_d ) \mathbf{x}_k-\mathbf{1}_n\otimes \mathbf{x}^{\ast} \rVert
\\
\label{Eq:A3}      \le& \lVert \mathbf{\tilde{s}}_{\text{w}}^{k} \rVert +nL\lVert \mathbf{\tilde{x}}_{\mathbf{w}}^{k} \rVert +nL\lVert \mathbf{r}^k \rVert. \tag{A3}
\end{flalign}
Since $\mathbf{s}^k=( ( \mathbf{\bar{V}}^k ) ^{-1}\otimes \mathbf{I}_d ) \mathbf{y}^k$, it holds $\mathbf{y}^k=( \mathbf{\bar{V}}^k\otimes \mathbf{I}_d ) \mathbf{s}^k$. Further, using the relation $\lVert \mathbf{\bar{V}}^k \rVert =\max _i[ \mathbf{v}_k ] _i<1$ gives
\begin{flalign}
\label{Eq:A4} \lVert \mathbf{y}^k \rVert \le \lVert \mathbf{\bar{V}}^k \rVert \lVert \mathbf{s}^k \rVert \le \lVert \mathbf{s}^k \rVert. \tag{A4}
\end{flalign}
Substituting the results of \eqref{Eq:A3} and \eqref{Eq:A4} into \eqref{Eq:A2} completes the proof.
\end{proof}

\section{Proof of Lemma 5}
\begin{proof}
It is easily verified from the definitions of $\mathbf{\bar{x}}_{\text{w}}^{k}$ and $\mathbf{r}^k$ that $\mathbf{r}^k=( \mathbf{1}_n( \boldsymbol{\phi }^k ) ^{\top}\otimes \mathbf{I}_d ) \mathbf{x}^k-\mathbf{1}_n\otimes \mathbf{x}^{\ast}$. Then, using the dynamic \eqref{Eq:10} gives
\begin{flalign}
\nonumber \mathbf{r}^{k+1}=&( \mathbf{1}_n( \boldsymbol{\phi }^{k+1} ) ^{\top}\otimes \mathbf{I}_d ) \mathbf{R}^k\mathbf{x}^k-\mathbf{1}_n\otimes \mathbf{x}^{\ast}
\\
\nonumber        &-\gamma ( \mathbf{1}_n( \boldsymbol{\phi }^{k+1} ) ^{\top}\otimes \mathbf{I}_d ) \mathbf{A}^k\mathbf{Ty}^k
\\
\nonumber    =&( \mathbf{1}_n( \boldsymbol{\phi }^k ) ^{\top}\otimes \mathbf{I}_d ) \mathbf{x}^k-\mathbf{1}_n\otimes \mathbf{x}^{\ast}
\\
\nonumber       &-\gamma [ ( \mathbf{1}_n( \boldsymbol{\phi }^{k+1} ) ^{\top}\mathbf{\bar{A}}^k\mathbf{\bar{T}v}^k\mathbf{1}_{\tilde{n}}^{\top} ) \otimes \mathbf{I}_d ] \mathbf{y}^k
\\
\nonumber       &-\gamma ( \mathbf{1}_n( \boldsymbol{\phi }^{k+1} ) ^{\top}\otimes \mathbf{I}_d ) \mathbf{A}^k\mathbf{T}( \mathbf{y}^k-( \mathbf{v}^k\mathbf{1}_{\tilde{n}}^{\top}\otimes \mathbf{I}_d ) \mathbf{y}^k ).
\end{flalign}
We bound $\mathbf{r}^{k+1}$ as
\begin{flalign}
\nonumber &\lVert \mathbf{r}^{k+1} \rVert
\\
\nonumber  \le &\left\| ( \mathbf{1}_n( \boldsymbol{\phi }^k ) ^{\top}\otimes \mathbf{I}_d ) \mathbf{x}^k-\mathbf{1}_n\otimes \mathbf{x}^{\ast} \right.
\\
\nonumber  &\,\, \left. -\gamma [ ( \mathbf{1}_n( \boldsymbol{\phi }^{k+1} ) ^{\top}\mathbf{\bar{A}}^k\mathbf{\bar{T}v}^k\mathbf{1}_{\tilde{n}}^{\top} ) \otimes \mathbf{I}_d ] \mathbf{y}^k \right\|
\\
\nonumber  &+\lVert \gamma ( \mathbf{1}_n( \boldsymbol{\phi }^{k+1} ) ^{\top}\otimes \mathbf{I}_d ) \mathbf{A}^k\mathbf{T}( \mathbf{y}^k-( \mathbf{v}^k\mathbf{1}_{\tilde{n}}^{\top}\otimes \mathbf{I}_d ) \mathbf{y}^k ) \rVert
\\
\nonumber \le &\lVert ( \mathbf{1}_n( \boldsymbol{\phi }^k ) ^{\top}\otimes \mathbf{I}_d ) \mathbf{x}^k-\mathbf{1}_n\otimes \mathbf{x}^{\ast}-\gamma \varrho ^k[ ( \mathbf{1}_n\mathbf{1}_{\tilde{n}}^{\top} ) \otimes \mathbf{I}_d ] \mathbf{y}^k \rVert
\\
\label{Eq:A5} &+\gamma n\lVert \mathbf{\tilde{s}}_{\text{w}}^{k} \rVert, \tag{A5}
\end{flalign}
where $\varrho ^k\coloneqq ( \boldsymbol{\phi }^{k+1} ) ^{\top}\mathbf{\bar{A}}^k\mathbf{\bar{T}v}^k\in \mathbb{R}$, and the last inequality uses the relations that $\lVert \mathbf{1}_n( \boldsymbol{\phi }^k ) ^{\top}\otimes \mathbf{I}_d \rVert \le \sqrt{n}$, $\lVert \mathbf{A}^k \rVert \le \sqrt{n}$, $\mathbf{y}^k-( \mathbf{v}^k\mathbf{1}_{\tilde{n}}^{\top}\otimes \mathbf{I}_d ) \mathbf{y}^k=( \mathbf{\bar{V}}^k\otimes \mathbf{I}_d ) \mathbf{\tilde{s}}_{\text{w}}^{k}$, and $\lVert \mathbf{\bar{V}}^k\otimes \mathbf{I}_d \rVert \le 1$. Note that $v_{i}^{k}\in [ \eta ^{\tilde{n}-1}/\tilde{n},1 ]$ for $k\ge 1$ \cite{Saadatniaki2020,Gao2023}. Using the row stochasticity of $\mathbf{\bar{A}}^k$, i.e., $\sum\nolimits_{j=1}^n{a_{ij}^{k}}=1$, yields $\eta ^{\tilde{n}-1}/\tilde{n}\le \sum\nolimits_{j=1}^n{a_{ij}^{k}v_{j}^{k}}\le 1$ for any $i=1,\cdots ,n$. Thus, it follows from the relation $\sum\nolimits_{i=1}^n{\phi _{i}^{k+1}}=1$ that
\begin{flalign}
\nonumber \varrho ^k=( \boldsymbol{\phi }^{k+1} ) ^{\top}\mathbf{\bar{A}}^k\mathbf{\bar{T}v}^k=\sum_{i=1}^n{\phi _{i}^{k+1}\sum_{j=1}^n{a_{ij}^{k}v_{j}^{k}}\in [ \eta ^{\tilde{n}-1}/\tilde{n},1 ]},
\end{flalign}
where the second equality uses the relation $[ \mathbf{\bar{T}v}^k ] _i=v_{i}^{k}$ for $i=1,\cdots ,n$. Besides, one can verify $( \mathbf{1}_{\tilde{n}}^{\top}\otimes \mathbf{I}_d ) \mathbf{y}^k=( \mathbf{1}_{\tilde{n}}^{\top}\otimes \mathbf{I}_d ) \nabla \mathbf{\tilde{f}}( \mathbf{x}^k ) =( \mathbf{1}_{n}^{\top}\otimes \mathbf{I}_d ) \nabla \mathbf{f}( \mathbf{x}^k )$. Thus, it holds
\begin{flalign}
\nonumber  &\lVert ( \mathbf{1}_n( \boldsymbol{\phi }^k ) ^{\top}\otimes \mathbf{I}_d ) \mathbf{x}^k-\mathbf{1}_n\otimes \mathbf{x}^{\ast}-\gamma \varrho ^k( \mathbf{1}_n\mathbf{1}_{\tilde{n}}^{\top}\otimes \mathbf{I}_d ) \mathbf{y}^k \rVert
\\
\nonumber =&\lVert \mathbf{1}_n\otimes \mathbf{\bar{x}}_{\text{w}}^{k}-\mathbf{1}_n\otimes \mathbf{x}^{\ast}-\gamma \varrho ^k( \mathbf{1}_n\mathbf{1}_{n}^{\top}\otimes \mathbf{I}_d ) \nabla \mathbf{f}( \mathbf{x}^k ) \rVert
\\
\nonumber \le &\lVert \mathbf{1}_n\otimes \mathbf{\bar{x}}_{\text{w}}^{k}-\mathbf{1}_n\otimes \mathbf{x}^{\ast}-\gamma \varrho ^k( \mathbf{1}_n\mathbf{1}_{n}^{\top}\otimes \mathbf{I}_d ) \nabla \mathbf{f}( \mathbf{1}_n\otimes \mathbf{\bar{x}}_{\text{w}}^{k} ) \rVert
\\
\nonumber &+\gamma \varrho ^k\lVert ( \mathbf{1}_n\mathbf{1}_{n}^{\top}\otimes \mathbf{I}_d ) ( \nabla \mathbf{f}( \mathbf{x}^k ) -\nabla \mathbf{f}( \mathbf{1}_n\otimes \mathbf{\bar{x}}_{\text{w}}^{k} ) ) \rVert
\\
\label{Eq:A6} \le &\sqrt{n}\lVert \mathbf{\bar{x}}_{\text{w}}^{k}-\mathbf{x}^{\ast}-\gamma \varrho ^k\nabla F( \mathbf{\bar{x}}_{\text{w}}^{k} ) \rVert +\gamma nL\lVert \mathbf{\tilde{x}}_{\text{w}}^{k} \rVert. \tag{A6}
\end{flalign}
Under Assumption 2, applying Lemma 10 in \cite{Qu2017} yields for $0<\gamma \le 1/\bar{L}$
\begin{flalign}
\label{Eq:A7} \lVert \mathbf{\bar{x}}_{\text{w}}^{k}\!-\!\mathbf{x}^{\ast}\!-\!\gamma \varrho ^k\nabla F( \mathbf{\bar{x}}_{\text{w}}^{k} ) \rVert \!\le\! \frac{1}{\sqrt{n}}( 1\!-\!\gamma \tilde{n}^{-1}\eta ^{\tilde{n}-1}\mu ) \lVert \mathbf{r}^k \rVert. \tag{A7}
\end{flalign}
Therefore, combining the relations \eqref{Eq:A5}, \eqref{Eq:A6}, and \eqref{Eq:A7} completes the proof.
\end{proof}

\section{Proof of Lemma 6}
\begin{proof}
Let $\mathbf{z}^k=\nabla \mathbf{\hat{f}}( \mathbf{x}^{k+1} ) -\nabla \mathbf{\hat{f}}( \mathbf{x}^k )$. Using the dynamic of \eqref{Eq:11} gives
\begin{flalign}
\nonumber \mathbf{s}^{k+1}=&\mathbf{P}_{k:k-\bar{N}+1}\mathbf{s}^{k-\bar{N}+1}+( ( \mathbf{\bar{V}}^{k+1} ) ^{-1}\otimes \mathbf{I}_d ) \mathbf{z}^k
\\
\nonumber\,\,      &+\sum_{l=1}^{\bar{N}-1}{\mathbf{P}_{k:k-l+1}( ( \mathbf{\bar{V}}^{k-l+1} ) ^{-1}\otimes \mathbf{I}_d ) \mathbf{z}^{k-l}}.
\end{flalign}
According to the definition of $\mathbf{\tilde{s}}_{\text{w}}^{k}$, one can verify that for $k\ge \bar{N}$
\begin{flalign}
\nonumber \lVert \mathbf{\tilde{s}}_{\text{w}}^{k+1} \rVert \le &r_P\lVert [ ( \mathbf{I}_{\tilde{n}}-\mathbf{1}_{\tilde{n}}( \mathbf{v}^{k-\bar{N}+1} ) ^{\top} ) \otimes \mathbf{I}_d ] \mathbf{s}^{k-\bar{N}+1} \rVert
\\
\nonumber \,\,          &+\lVert \mathbf{I}_{\tilde{n}}-\mathbf{1}_{\tilde{n}}( \mathbf{v}^{k+1} ) ^{\top} \rVert \lVert ( \mathbf{\bar{V}}^{k+1} ) ^{-1} \rVert \lVert \mathbf{z}^k \rVert
\\
\nonumber \,\,          &+\sum_{l=1}^{\bar{N}-1}{\lVert \mathbf{I}_{\tilde{n}}-\mathbf{1}_{\tilde{n}}( \mathbf{v}^{k-l+1} ) ^{\top} \rVert \lVert ( \mathbf{\bar{V}}^{k-l+1} ) ^{-1} \rVert \lVert \mathbf{z}^{k-l} \rVert}
\\
\label{Eq:A8}\,\,       \le &r_P\lVert \mathbf{\tilde{s}}_{\text{w}}^{k-\bar{N}+1} \rVert +\tilde{n}Q_P/\eta ^{\tilde{n}-1}\sum_{l=0}^{\bar{N}-1}{\lVert \mathbf{z}^{k-l} \rVert}, \tag{A8}
\end{flalign}
where the last inequality uses the relations $\lVert \mathbf{I}_{\tilde{n}}-\mathbf{1}_{\tilde{n}}( \mathbf{v}^{k+1} ) ^{\top} \rVert \le 2\sqrt{\tilde{n}}\le Q_P$, $\lVert ( \mathbf{\bar{V}}^k ) ^{-1} \rVert \le \tilde{n}/\eta ^{\tilde{n}-1}$, and the result in Lemma 2.

From the dynamic of \eqref{Eq:10}, we can bound $\mathbf{z}^k$ as
\begin{flalign}
\nonumber \lVert \mathbf{z}^k \rVert \le &\Big( \sum_{i=1}^n{L_{i}^{2}\lVert \mathbf{x}_{i}^{k+1}-\mathbf{x}_{i}^{k} \rVert ^2} \Big) ^{1/2}\le L\lVert \mathbf{x}^{k+1}-\mathbf{x}^k \rVert
\\
\nonumber     \le &L\lVert ( \mathbf{R}^k-\mathbf{I}_{nd} ) \mathbf{x}^k \rVert +\gamma L\lVert \mathbf{A}^k\mathbf{Ty}^k \rVert
\\
\nonumber     \le &L\lVert ( \mathbf{R}^k\!-\!\mathbf{I}_{nd} ) [ ( \mathbf{I}_n\!-\!\mathbf{1}_n( \boldsymbol{\phi }^k ) ^{\top} ) \!\otimes\! \mathbf{I}_d ] \mathbf{x}^k \rVert \!+\!\gamma L\lVert \mathbf{A}^k\mathbf{Ty}^k \rVert
\\
\nonumber     \le &2\sqrt{n}L\lVert \mathbf{\tilde{x}}_{\text{w}}^{k} \rVert +\gamma L\sqrt{n}\lVert \mathbf{y}^k \rVert
\\
\nonumber \le &( 2\sqrt{n}L+\gamma n\sqrt{n}L^2 ) \lVert \mathbf{\tilde{x}}_{\text{w}}^{k} \rVert
\\
\label{Eq:A9} &+\gamma n\sqrt{n}L^2\lVert \mathbf{r}^k \rVert +\gamma \sqrt{n}L\lVert \mathbf{\tilde{s}}_{\text{w}}^{k} \rVert, \tag{A9}
\end{flalign}
where the last second inequality uses the facts that $\lVert \mathbf{R}^k-\mathbf{I}_{nd} \rVert \le 2\sqrt{n}$ and $[ ( \mathbf{I}_n-\mathbf{1}_n( \boldsymbol{\phi }^k ) ^{\top} ) \otimes \mathbf{I}_d ] \mathbf{x}^k=\mathbf{\tilde{x}}_{\text{w}}^{k}$, as well as the last inequality uses the results in \eqref{Eq:A3} and \eqref{Eq:A4}.

Substituting \eqref{Eq:A9} into \eqref{Eq:A8} yields the desired result.
\end{proof}

\vfill

\end{document}